\documentclass[aps,12pt,axodraw,nofootinbib,superscriptaddress,aps]{revtex4}
\pdfoutput=1
\usepackage{epsfig}
\usepackage{amsmath,amssymb}
\usepackage{bm}
\usepackage{times}
\usepackage{graphicx}
\usepackage{epstopdf}
\usepackage{amsfonts}
\usepackage{bm}
\usepackage{epsfig}
\usepackage{graphics}
\usepackage{xspace}
\usepackage[usenames]{color}
\usepackage{multirow}

\def\ba{\begin{eqnarray}}
\def\ea{\end{eqnarray}}
\def\be{\begin{equation}}
\def\ee{\end{equation}}

\def\beq{\begin{equation}}
\def\eeq{\end{equation}}
\newcommand{\slashed}{\slash \hspace{-0.23cm}}

 \def\bea{\begin{eqnarray}}
\def\eea{\end{eqnarray}}
\newcommand{\nbea}{\begin{align*}}
\newcommand{\neea}{\end{align*}}
\newcommand{\nbeq}{\begin{equation*}}
\newcommand{\neeq}{\end{equation*}}

\numberwithin{equation}{section}

\begin{document}

\title{\Large A Fast Track towards the `Higgs' Spin and Parity\\}

\vspace{2cm}

\author{John Ellis}
\affiliation{Theoretical Particle Physics and Cosmology Group, Physics Department, King's College London, Strand, London, UK}
\affiliation{Theory Division, Physics Department, CERN, CH-1211 Geneva 23, Switzerland}

\author{Dae Sung Hwang}
\affiliation{Department of Physics, Sejong University, Seoul 143Ð747, South Korea}

\author{Ver\'onica Sanz}
\affiliation{Theory Division, Physics Department, CERN, CH-1211 Geneva 23, Switzerland}
\affiliation{Department of Physics and Astronomy, York University, Toronto, ON, Canada, M3J 1P3}

\author{Tevong You}
\affiliation{Theoretical Particle Physics and Cosmology Group, Physics Department, KingÕs College London, Strand, London, UK}
\affiliation{Theory Division, Physics Department, CERN, CH-1211 Geneva 23, Switzerland}

%%%%%%%%%%%%%%%%%%%%%%%%%%%%%%%%%%%%%%%%%%%%%%%%%%%%%%%%%%%%%%%%%%%%
\date{\today}
%%%%%%%%%%%%%%%%%%%%%%%%%%%%%%%%%%%%%%%%%%%%%%%%%%%%%%%%%%%%%%%%%%%%

\vspace{2cm}

\begin{abstract}

\begin{centering}
~\\
{\bf Abstract} \\
~\\
\end{centering}

The LHC experiments ATLAS and CMS have discovered a new boson that resembles the
long-sought Higgs boson: it cannot have spin one, and has couplings to other particles
that increase with their masses, but the spin and parity remain to be determined.
We show here that the `Higgs' + gauge boson invariant-mass distribution
in `Higgs'-strahlung events at the Tevatron or the LHC would be very different under the
$J^P = 0^+, 0^-$ and $2^+$ hypotheses, and could provide a fast-track indicator of
the `Higgs' spin and parity. Our analysis is based on simulations of the experimental
event selections and cuts using {\tt PYTHIA} and {\tt Delphes}, and incorporates
statistical samples of  `toy' experiments. \\

~~~~~~~~~~~~~~~~~KCL-PH-TH/2012-38, LCTS/2012-22, CERN-PH-TH/2012-226

\end{abstract}
%%%%%%%%%%%%%%%%%%%%%%%%%%%%%%%%%%%%%%%%%%%%%%%%%%%%%%%%%%%%%%%%%%%%
\maketitle
%%%%%%%%%%%%%%%%%%%%%%%

%%%%%%%%%%%%%%%%%%%%%%%%%%%%%%%%%%%%%%%%%%%%%%%%%%%%%%%%%%%%%%%%%%%%
\section{Introduction}
%%%%%%%%%%%%%%%%%%%%%%%%%%%%%%%%%%%%%%%%%%%%%%%%%%%%%%%%%%%%%%%%%%%%

The new particle $X$ with mass $\sim 125$ to 126~GeV that has been discovered by the LHC experiments
ATLAS~\cite{ATLAS} and CMS~\cite{CMS}, with support from the TeVatron experiments
CDF and D0~\cite{CDFD0}, has similarities to the long-sought Higgs particle $H$.
The $X$ particle is a boson that does not have spin one, and its couplings to other particles depend
on their masses in a way very similar to the linear dependence expected for the Higgs boson of the Standard Model~\cite{EY2}.
However, the spin and parity $J^P$ of the $X$ particle remain to be determined, and this should
be regarded as an open question, with the pseudoscalar hypothesis $J^P = 0^-$ and the
tensor hypothesis $J^P = 2^+$ being important possibilities to exclude.

Various strategies have been proposed for determining the spin and parity of a Higgs candidate
in hadron-hadron collisions, including angular distributions and kinematic correlations in 
$X \to Z Z^\ast, W W^\ast$ and $\gamma \gamma$ decays~\cite{hspin}. Historically, the problem of
determining the spin and parity of a Higgs candidate was first considered in the context of
$e^+ e^-$ collisions, and the point was made that the threshold behaviour of the
cross section for the `Higgs'-strahlung process $e^+ e^- \to Z + X$ would depend on the
spin and parity of the $X$ particle, offering potential discrimination between different
spin-parity assignments~\cite{MCEMZ}.

In this paper we point out that calculations of the $V + X$ invariant mass distributions in antiproton-proton
collisions at the Tevatron collider and proton-proton collisions at the LHC reflect these differences in threshold behaviour.
In particular, the mean invariant mass $\langle M_{VX} \rangle$, as calculated using {\tt HELAS}~\cite{helas-spin2} 
and {\tt MadGraph}~\cite{MG5}, would be very different in the 
$J^P (X) = 0^+, 0^-$ and $2^+$ cases, where we assume graviton-like couplings in the latter case.
Specifically, we find in both parton-level simulations using {\tt PYTHIA}~\cite{PYTHIA} and more detailed
detector simulations using {\tt Delphes}~\cite{Delphes} that $\langle M_{VX(0^+)} \rangle \ll \langle M_{VX(0^-)} \rangle
\ll \langle M_{VX(2^+)} \rangle$, also after applying the experimental event selections and cuts. 
We use statistical samples of  `toy' experiments to analyze the potential
discriminating power of the TeVatron and LHC experiments. These demonstrate that they may (soon) be able to
discriminate between different $J^P$ assignments for the $X$ particle using the
$V + X$ invariant mass distribution, which could provide a `fast track' towards determining
its spin and parity.

%%%%%%%%%%%%%%%%%%%%%%%%%%%%%%%%%%%%%%%%%%%%%%%%%%%%%%%%%%%%%%%%%%%%
\section{Calculations for different spin-parity assignments}
%%%%%%%%%%%%%%%%%%%%%%%%%%%%%%%%%%%%%%%%%%%%%%%%%%%%%%%%%%%%%%%%%%%%

The fact that the $X$ particle has been observed to decay into a pair of on-shell photons
implies, as is well known, that it cannot have spin one. The simplest possibilities are that
it has spin zero or spin two, both of which occur in some theoretical frameworks. For example, there are
many proposals for particles with the pseudoscalar assignment $J^P = 0^-$, as well as the 
assignment $0^+$ expected for the Higgs boson of the Standard Model, and models
postulating extra dimensions raise the possibility of a massive spin-two particle.

In the case of the $0^+$ assignment for the $X$ particle, we assume the minimal $V_\mu V^\mu X$
coupling, and in the $0^-$ case we assume the dimension-five effective coupling 
$\epsilon_{\mu \nu \rho \sigma} F^{\mu \nu} F^{\rho \sigma} X$, where $F_{\mu \nu}$ is the
field-strength tensor of the vector boson $V$. In the case of a spin-two $X$ particle, there
is considerable ambiguity in the possible couplings, with a five-parameter set of possibilities
considered in~\cite{MCEMZ} for the $2^+$ assignment, and a set of four possibilities for the
$2^-$ case. We study the option that we consider the best motivated, namely the $2^+$
assignment with graviton-like couplings to all other particles including vector bosons~\footnote{We
note that Lorentz invariance and Standard Model gauge symmetries forbid dimension-four
couplings of a massive spin-two particle, and that the flavour and CP symmetries of the Standard Model
require it to couple flavour-diagonally to other particles via dimension-five terms that take the same forms as their
energy-momentum tensors~\cite{us-G}.}.
We use in our simulations the {\tt HELAS} library~\cite{helas-spin2}, including its implementation of a massive
spin-two particle with graviton-like couplings, and generate its production and decays using
{\tt MadGraph}. We implemented the pseudo-scalar couplings to gauge bosons and fermions, as well as 
{\tt Feynrules}~\cite{Feynrules} and the UFO model format~\cite{UFO} for implementation into {\tt Madgraph}.

As already mentioned, the reaction $e^+ e^- \to Z + X$ was shown in~\cite{MCEMZ} to exhibit
significant differences in the energy dependence of the total cross section for $X$ production in 
the $J^P = 0^{+,-}, 2^+$ cases, and other possible $J^P$ assignments were also considered.
Here we apply the considerations of~\cite{MCEMZ} to the related processes ${\bar p} p, pp \to \{Z, W\} + X$.
Discriminating power is provided by the threshold behaviour of the cross section. We recall that
production is an s-wave process in the $0^+$ case, so that the cross section rises $\sim \beta$ 
close to threshold. In the $0^-$ case, on the other hand, the production mechanism is p-wave, 
and the threshold behaviour $\sim \beta^3$. In the $2^+$ case, many of the possible couplings
make d-wave contributions to $V + X$ production amplitudes, yielding contributions to the total
cross section $\sim \beta^5$, and these contributions dominate in the case of graviton-like couplings.

\begin{figure}[h!]
\centering
\includegraphics[scale=0.38]{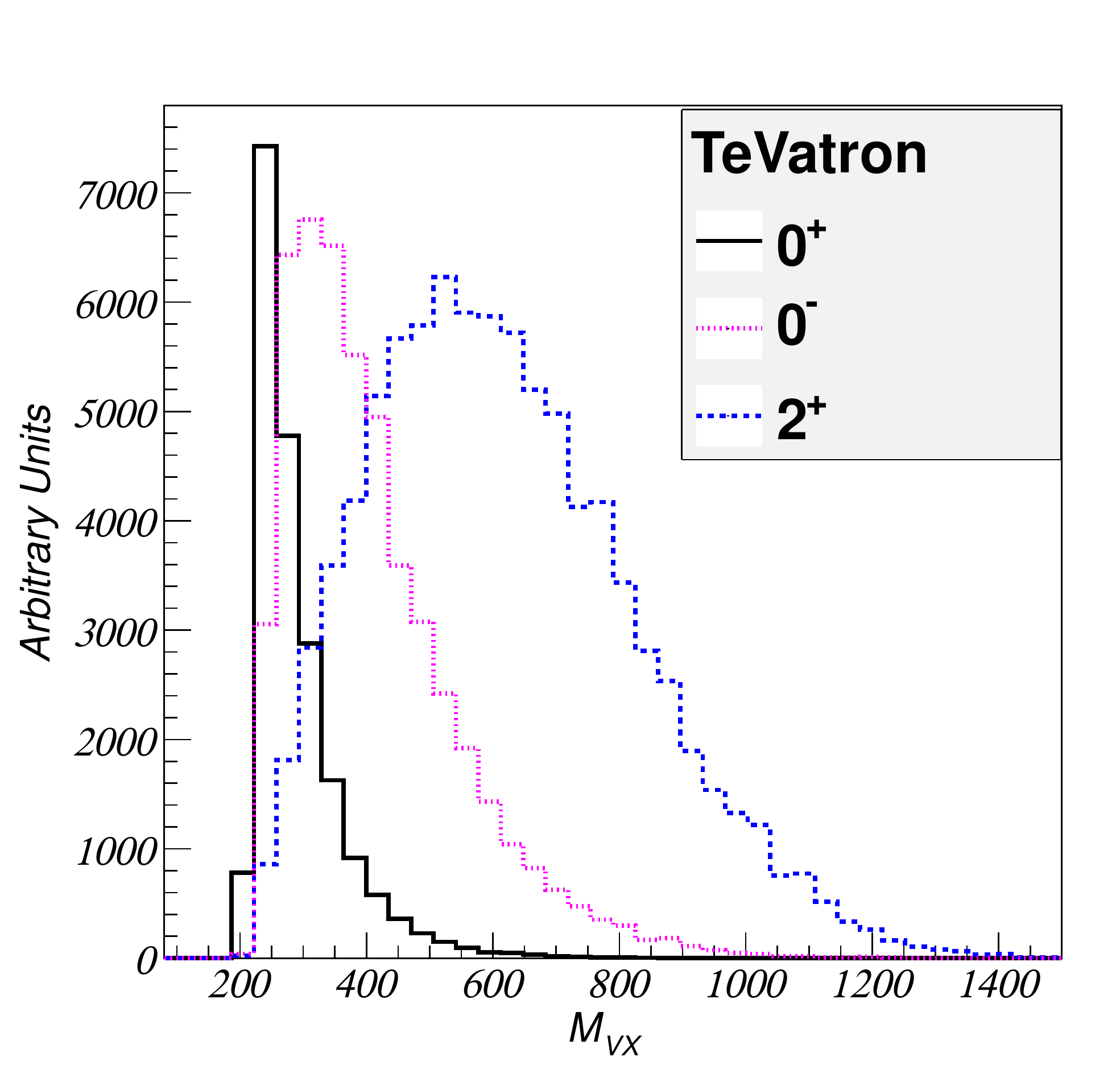}
\includegraphics[scale=0.38]{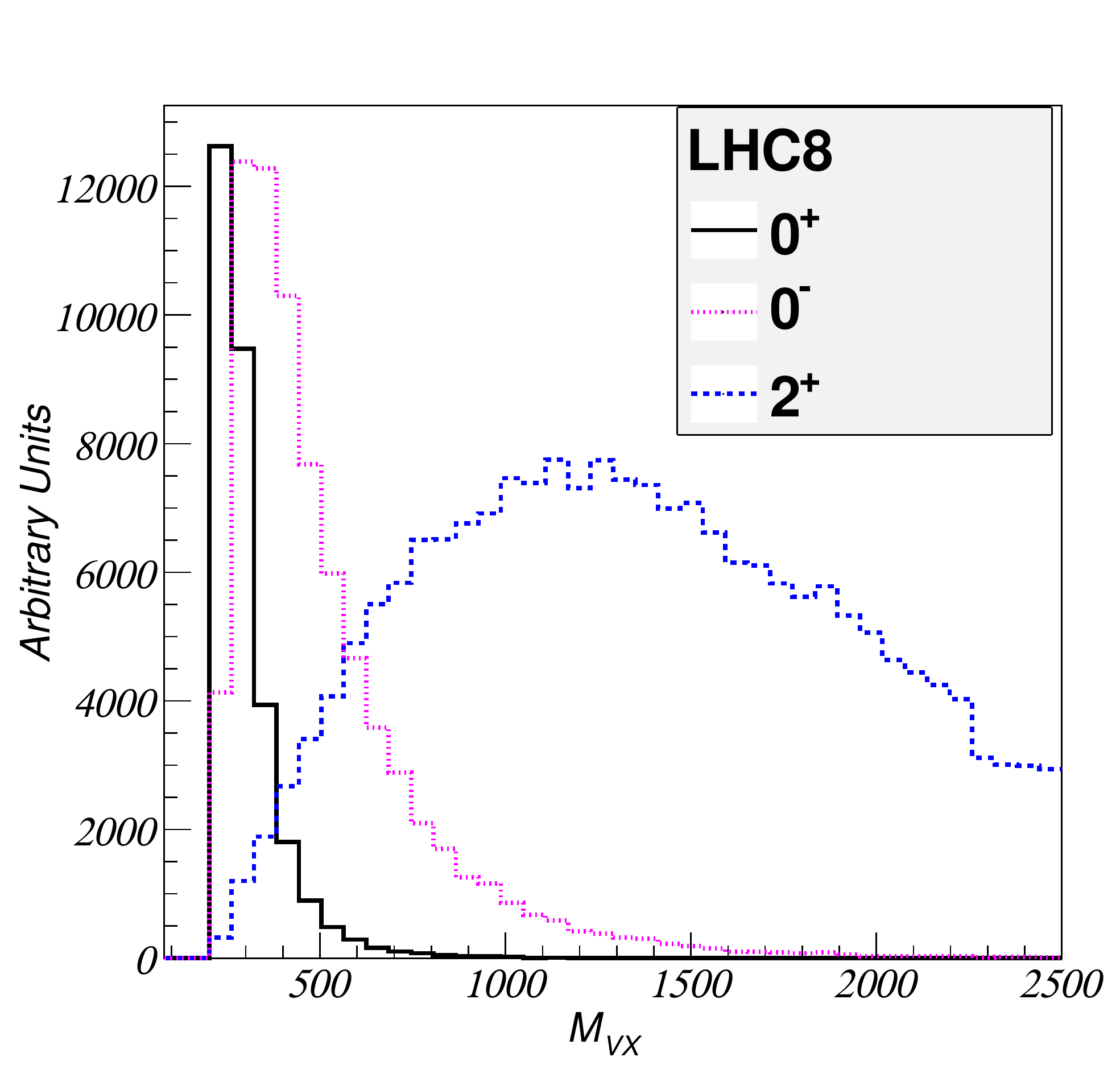}
\caption{\it The distributions in the $Z + X$ invariant mass $M_{ZX}$ for the $0^{+}$ (solid black), 
$0^{-}$ (pink dotted) and $2^+$ (blue dashed) assignments for the particle $X$ with mass $\sim 125$~GeV 
discovered by ATLAS~\cite{ATLAS} and CMS~\cite{CMS}, calculated for the reaction ${\bar p} p \to Z + X$ at the TeVatron
(left) and for the reaction $p p \to Z + X$ at the LHC at 8 TeV (right).}
\label{fig:beta1or3}
\end{figure}

\begin{figure}[h!]
\centering
\includegraphics[scale=0.38]{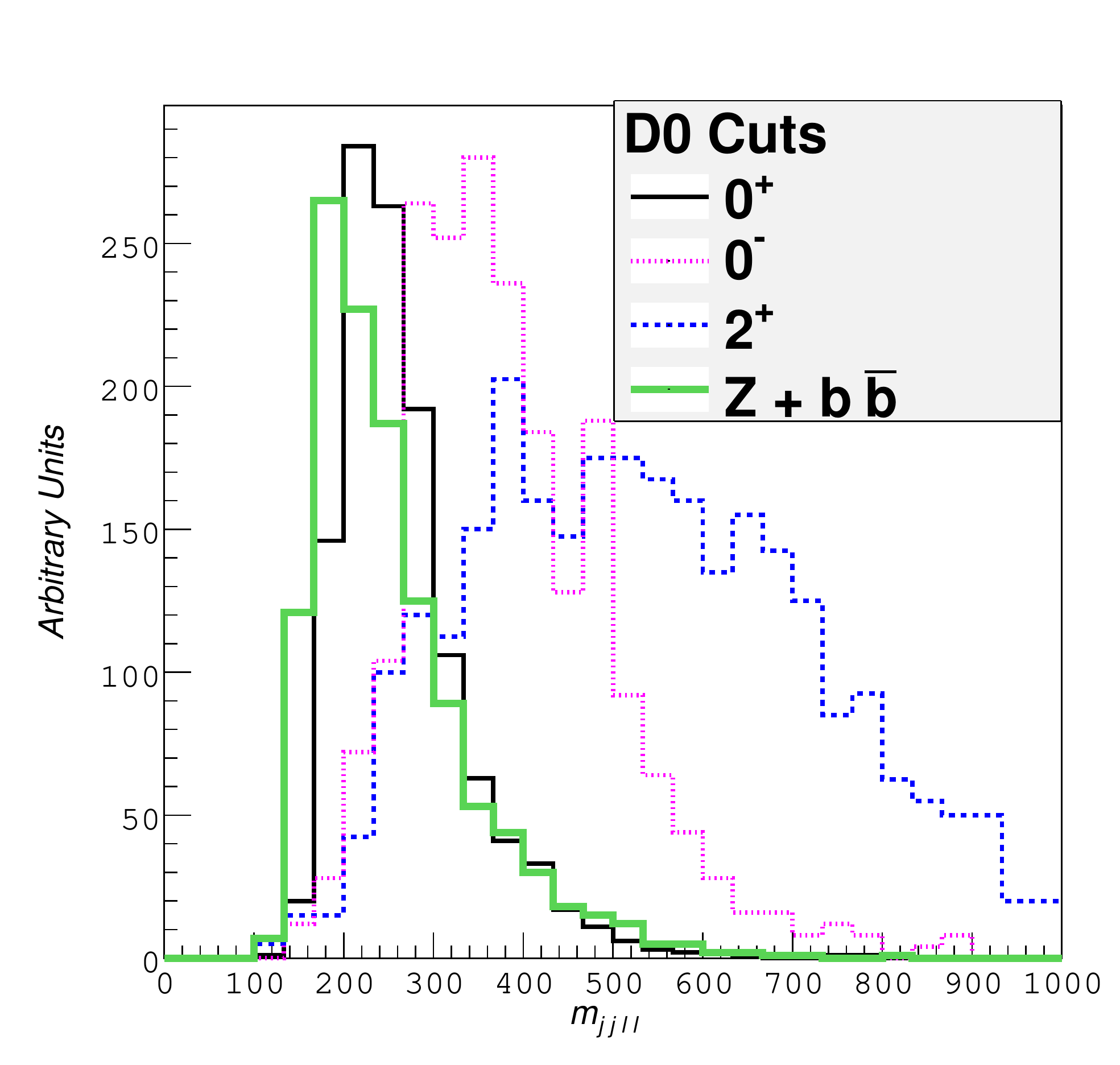}
\includegraphics[scale=0.38]{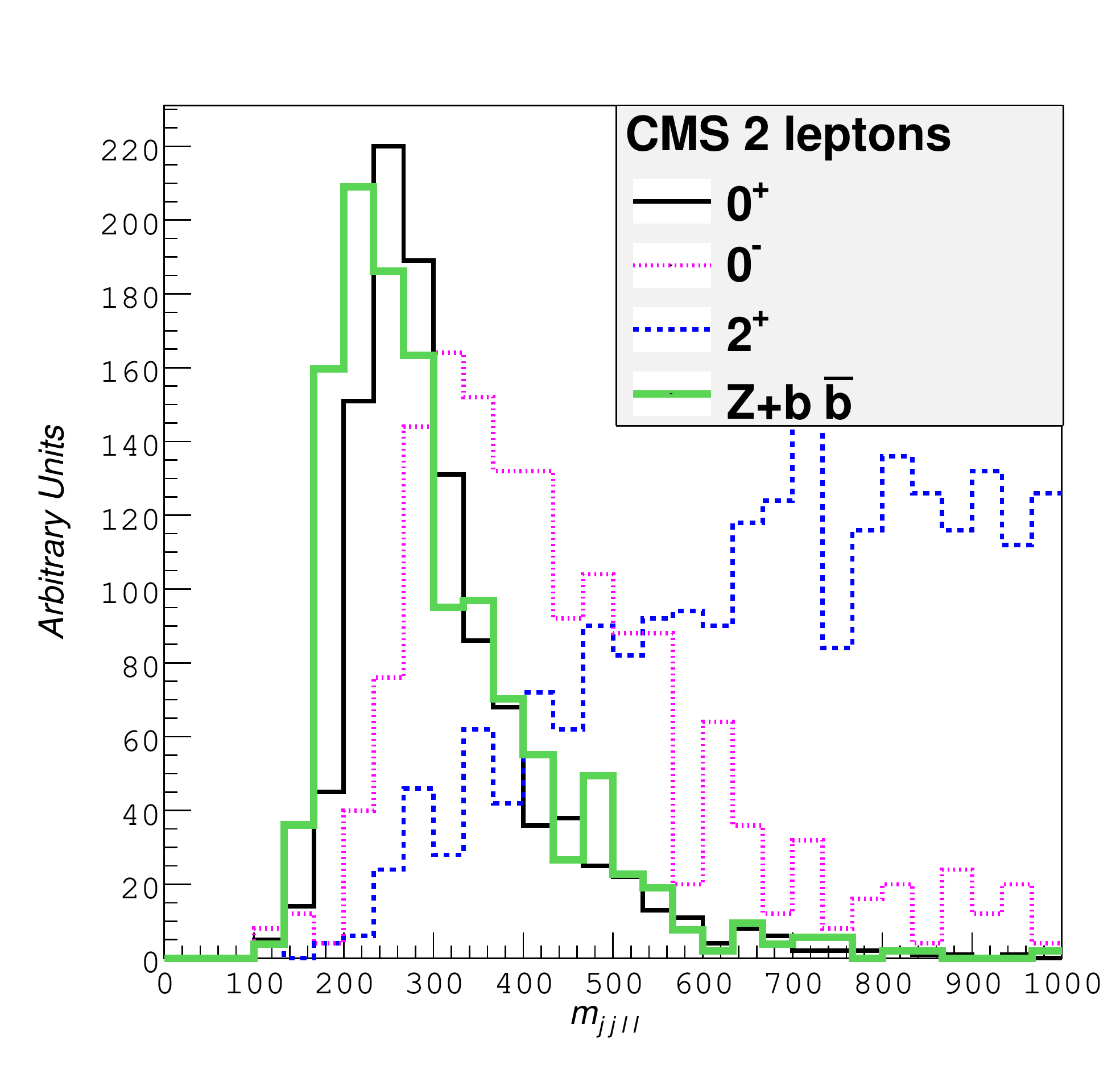}
\caption{\it The $Z + {\bar b} b$ background invariant mass distribution (green) at the TeVatron 
using the D0 cuts described in the text (left panel)
and the LHC at 8~TeV using the CMS cuts also described in the text
(right panel) compared with the two-lepton signal distributions in the $Z + X$ invariant mass $M_{ZX}$ for the $0^{+}$ (solid black), 
$0^{-}$ (pink dotted) and $2^+$ (blue dashed) assignments for the particle $X$ with mass $\sim 125$~GeV.}
\label{fig:BG}
\end{figure}

Fig.~\ref{fig:beta1or3} compares the (arbitrarily normalized) $Z + X$
invariant mass ($M_{ZX}$) distributions in the $J^P = 0^+$ case (solid black lines),
the $J^P = 0^-$ case (dotted pink lines) and the graviton-like $J^P = 2^+$ case 
(dashed blue line ) cases at the TeVatron (left panel) and at the LHC at 8 TeV (right panel), 
as simulated using {\tt MadGraph}~\cite{MG5} and {\tt PYTHIA}~\cite{PYTHIA}
at the parton level without including any detector simulation.
The results for the different spin-parity assignments are clearly very different, 
yielding large differences in the mean values of $\langle M_{ZX} \rangle$.
At the parton level, we find the following values for the distances above threshold, 
$\langle M_{ZX} \rangle - M_Z- M_X$, in the $J^P = 0^+, 0^-$ and $2^+$ cases: 
\bea
& \; \; \; \; \; \; \; {\rm TeVatron} & \; \; \; {\rm LHC~at~8~TeV} \nonumber \\
 & \, (0^+) \; \; ~~75 \textrm{ GeV}; & \; \; \; 88 \textrm{ GeV } \nonumber \\
\langle M_{ZX} \rangle - M_Z- M_X \; = \; & \; (0^-) \; \; 194  \textrm{ GeV}; & \;  \; \; 303  \textrm{ GeV} \nonumber \\
& \, (2^+) \; \; 400~  \textrm{GeV}; & \; \; \; 1340  \textrm{ GeV }\; .
\label{thresholds}
\eea
For comparison, we note that the invariant mass distributions for the $Z + {\bar b} b$ background,
shown as the green histograms in Fig.~\ref{fig:BG} for the TeVatron 
using the D0 cuts described below (left panel) and the LHC at 8~TeV
using the CMS cuts also described below (right panel),
are sharply peaked towards low invariant masses close to threshold, even closer than the $J^P = 0^+$
case (\ref{thresholds}).

Encouraged by the differences seen in (\ref{thresholds}) and in Fig.~\ref{fig:beta1or3}, we have made simulations of the
possible signals in the TeVatron and LHC experiments. We have not analyzed further the backgrounds in the
experiments, which would require more extensive simulations beyond the scope of this work.

\section{Detector simulations for different spin-parity assignments}

\subsection{TeVatron}

\begin{figure}[h!]
\centering
\includegraphics[scale=0.27]{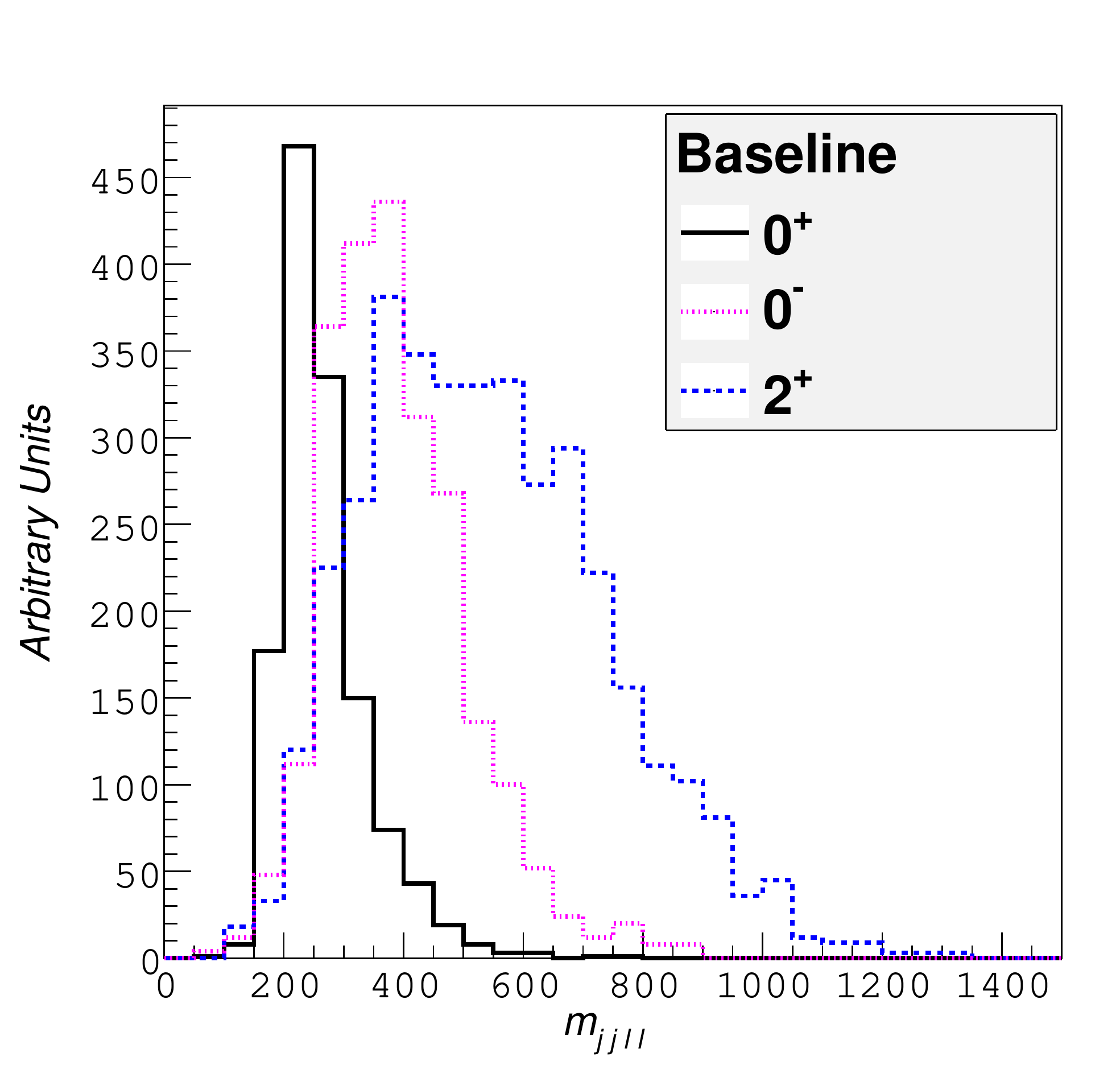}
\includegraphics[scale=0.27]{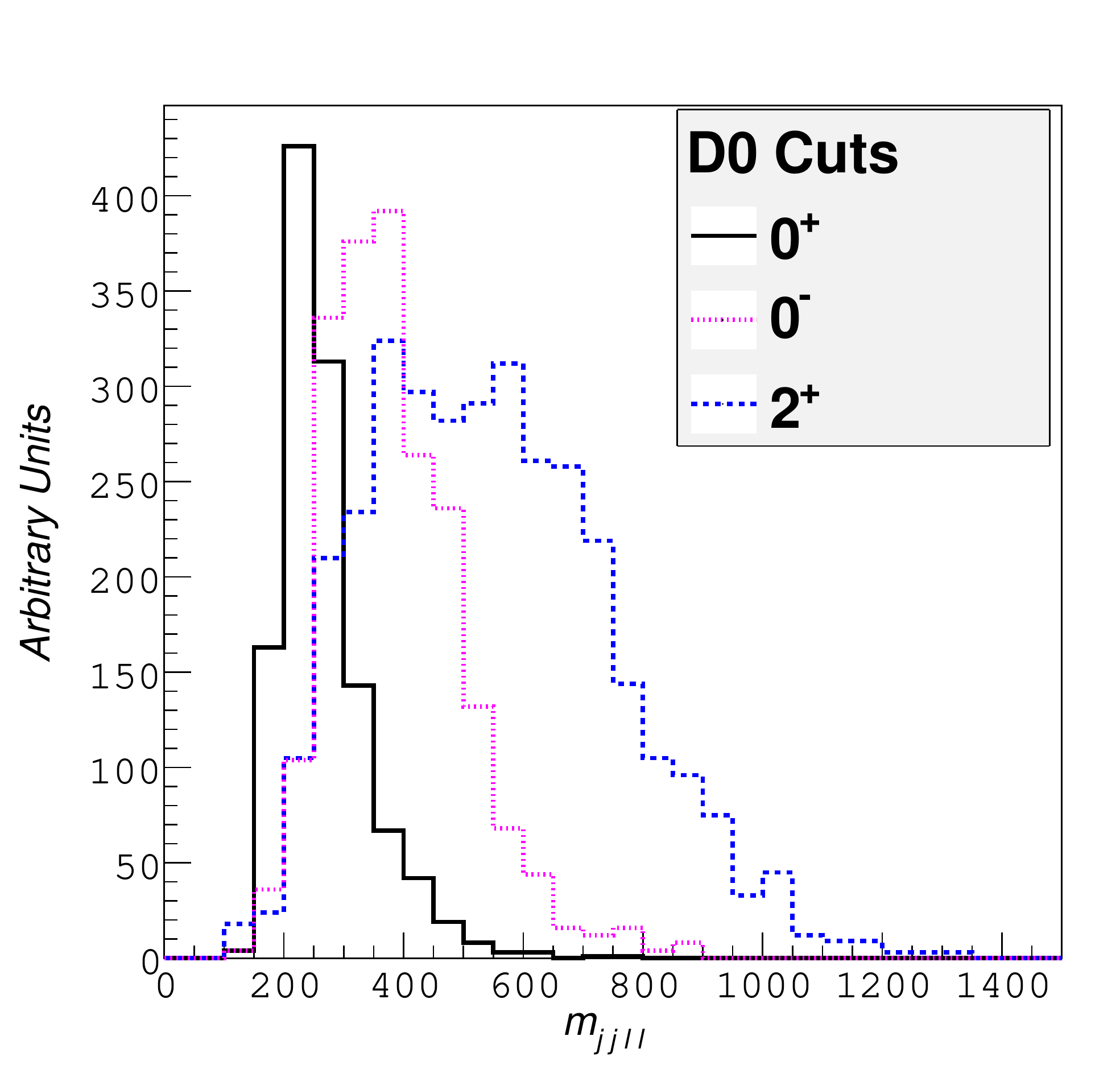}
\includegraphics[scale=0.27]{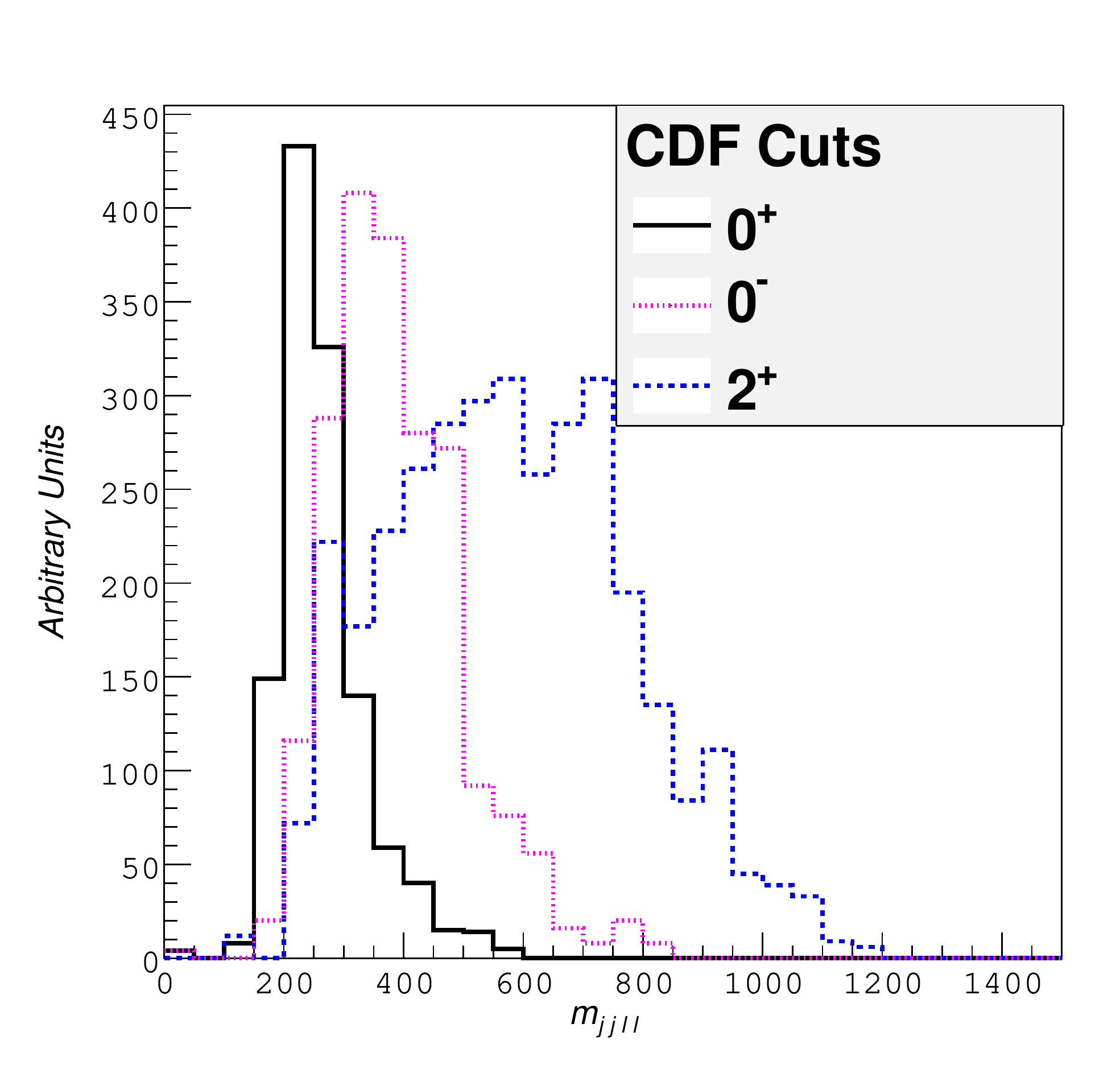}
\caption{\it The effect of fast simulations of $Z \to \ell^+ \ell^- + X \to \bar{b} b$ analyses
with {\tt Delphes}, using baseline (left panel), D0~\cite{D0-2l} (centre panel)
and CDF~\cite{CDF-2l} cuts (right panel). The discrimination between different $J^P$ assignments seen in
Fig.~\ref{fig:beta1or3} is maintained.}
\label{2l-cuts}
\end{figure}

The TeVatron experiments CDF and D0 have reported evidence for production of the $X$ particle in
association with $Z \to \ell^+ \ell^-, {\bar \nu} \nu$ and $W^\pm \to \ell^\pm \nu$~\cite{CDFD0}. In this Section we simulate
these analyses using {\tt Delphes}. We first apply the following baseline
parton-level cuts at the generator level: $p_T^{\ell} >$ 10 GeV, $|\eta_{\ell}|< 2.$, 
$p_T^{j} >$ 20 GeV, $|\eta_j|<$2.5 and $\Delta R_{j \ell}>$  0.5, where $\eta$ is the pseudo-rapidity
and $R$ is the standard cone angle variable, and jets are reconstructed using the cone size $R=0.5$
As shown in the left panel of Fig.~\ref{2l-cuts}, the discrimination between the different possible
spin-parity assignments survives the baseline cuts. We next proceed to implement event selections and cuts
specific to the CDF and D0 experiments for analyses with two, one and zero leptons.

\subsubsection{D0 and CDF $Z \to \ell^+ \ell^- + X \to \bar{b} b$ analyses}

The D0~\cite{D0-2l} selection cuts we implement are different for muons and electrons. In the muon case, 
we ask for a leading lepton with $p_T>$20 GeV and $|\eta_{\ell}|<$  2, and a  sub-leading lepton 
with $p_T> $ 15 GeV and $|\eta_{\ell}|<$  1.5. The electron category is characterized by two leptons 
of which at least one has $p_T>$ 15 GeV, and $|\eta_{\ell}|< $ 1.1. Summing over the single-
and double-tag categories, and adding the errors in quadrature, the number of events is $5.4\pm 0.3$. 
The CDF~\cite{CDF-2l} cuts we implement are more stringent. We ask for two or three jets, of which
at least two have $p_T> $ 25 GeV, $|\eta_j|< $2.5, and $m_{jj}> $ 25 GeV. 
Summing over all $b$-tagging categories, and adding the errors in quadrature, the number of events is 
$7.2\pm 0.6$. Making fast simulations using {\tt Delphes}, we find that the effects of these cuts on the $M_{ZX}$
distributions are mild, so that the discrimination between the different quantum numbers
assignments is maintained, as shown in the centre and right panels of Fig.~\ref{2l-cuts},
for the D0 and CDF experiments, respectively.

\begin{figure}[h!]
\centering
\includegraphics[scale=0.38]{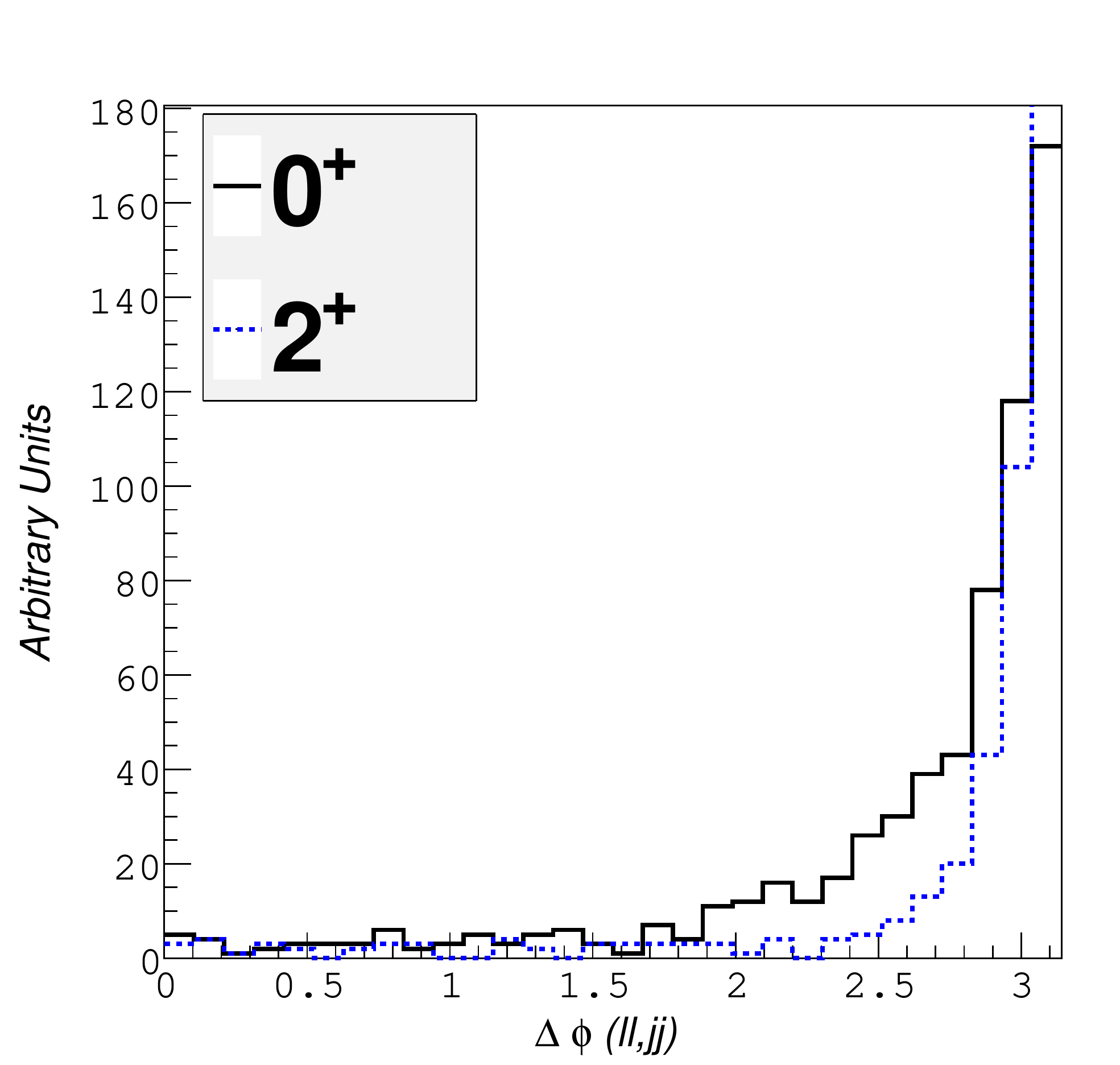}
\caption{\it The distributions in the difference in azimuthal angles, $\Delta \phi$, 
between the dijet and dilepton systems  in the TeVatron $Z \to \ell^+ \ell^- + X \to \bar{b} b$ analyses, 
for the $0^+$ and $2^+$ spin assignments (black solid
and blue dashed lines, respectively).}
\label{RF}
\end{figure}

After the selection cuts, both collaborations perform a multivariate analysis. 
One can find in~\cite{D0-2l}  a list of the variables used to train the random forest analysis, 
whose distribution depends on the quantum numbers of the candidate. The training was 
optimized for the $0^+$ hypothesis, and this could impact the overall efficiency of the analysis in
the cases of the $0^-$ and $2^+$ assignments. To illustrate this point, we show in Fig.~\ref{RF} the 
differences in the distribution of the difference in azimuthal angles, $\Delta \phi$, 
between the dijet and dilepton systems for the $0^+$ and $2^+$ spin assignments. 
However, the main discriminating variable is the dijet invariant mass, which is the same for all cases,
so we expect only a moderate effect from the sensitivity of the angular variables to the $X$ quantum numbers.

\subsubsection{CDF $W^\pm \to \ell^\pm \nu + X \to \bar{b}b$ analysis}

\begin{figure}[h!]
\centering
\includegraphics[scale=0.38]{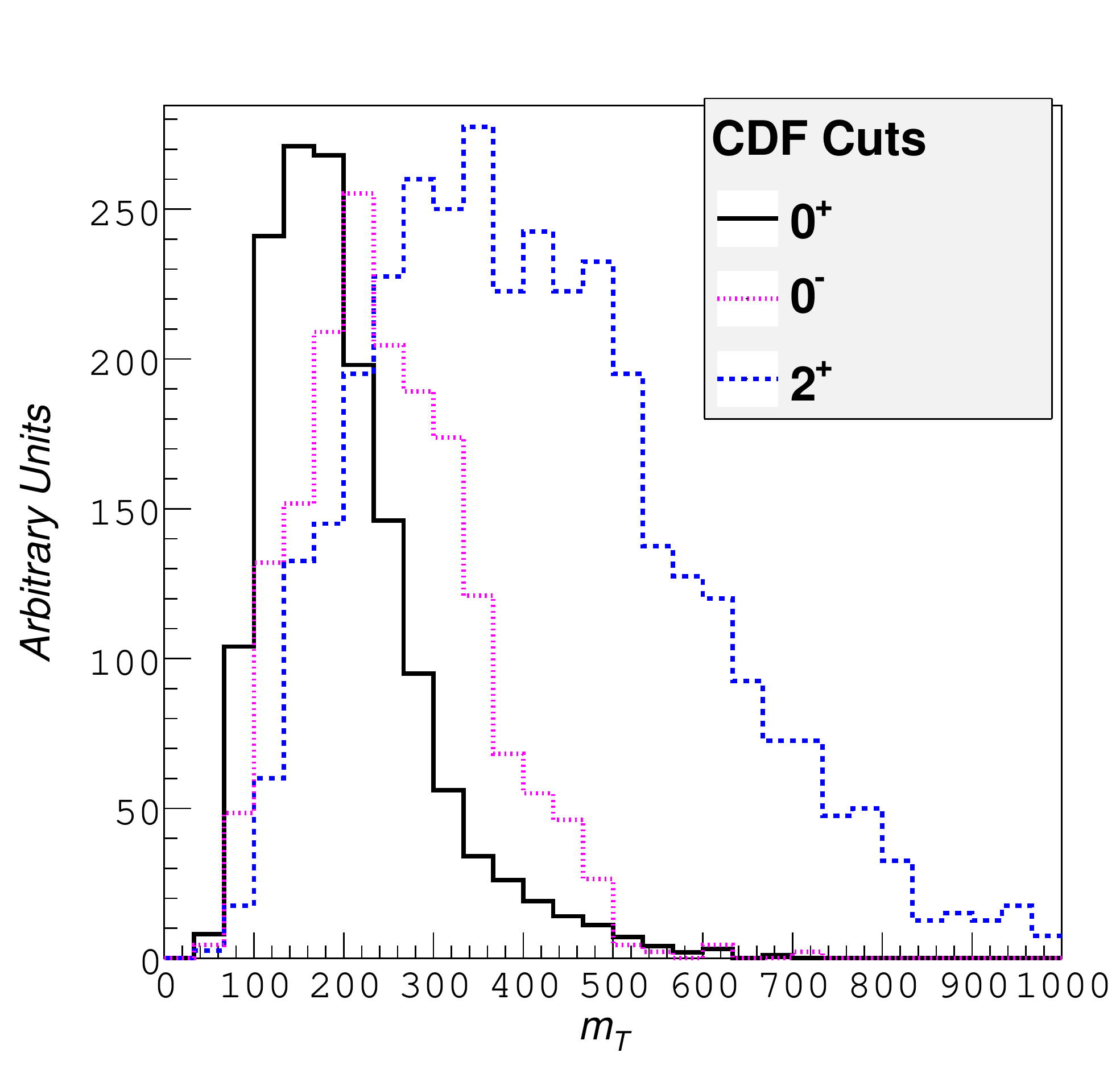}
\caption{\it The effect of a fast simulation of the CDF $W^\pm \to \ell^\pm \nu + X \to \bar{b}b$ analysis~\cite{CDF-1l}
with {\tt Delphes} on the transverse mass distributions for the different $J^P$ assignments.
The discrimination seen in Fig.~\ref{fig:beta1or3} is maintained.}
\label{CDF-1l}
\end{figure}

The analysis of this single-lepton channel has been published by CDF~\cite{CDF-1l}, and
$25.3\pm 1.4$ signal events were expected In the four two-jet categories.
At the parton level, we impose the cuts $p_T^{\ell} >$ 20 GeV, $|\eta_{\ell}|< 2.5$, 
$p_T^{j} >$ 20 GeV, $|\eta_j|<$2.5 and $\Delta R_{j \ell}>$  0.5.
The CDF analysis requires exactly two or three jets with $p_T^j>$ 20 GeV and $|\eta_j|<$2.  
There is also a cut on missing energy that depends on the centrality of the lepton, 
with tighter cuts for forward leptons. If $|\eta_{\ell}|<$ 1.1, the missing transverse
energy $\slashed{E}_T$ is required to be above 20~GeV, 
increases to 25~GeV in the forward region.
The selection cuts maintain the discrimination between different $J^P$ assignments
in the transverse mass variable
\bea
m_T^2=(E_T^W+E_T^X)^2- (\vec{p}_T^W+\vec{p}_T^X)^2 ,
\eea
where the W transverse momentum is
\bea
\vec{p}_T^W=\slashed{\vec{E}}_T+\vec{p}_T^{\ell} ,
\eea
as is shown in Fig.~\ref{CDF-1l}.  

\subsubsection{CDF and D0 $Z \to {\bar \nu} \nu + X \to \bar{b} b$ analyses}

Both CDF~\cite{CDF-0l} and D0~\cite{D0-0l} have published analyses of events with $\slashed{E}_T$
and no detected leptons, which are sensitive to $X$ production in association with $Z\to \nu \bar{\nu}$. 
We use {\tt Delphes} to reproduce the selection cuts in both analysis.
In the case of D0,  the relevant cuts are $p_T^j> $ 20 GeV, $|\eta_j|<$ 2.5, $\Delta \phi_{j_1j_2}<$  165$^0$, 
$\slashed{E}_T > $ 40 GeV, $H_T= |p_T^{j_1}|+ |p_T^{j_2}|>$  80 GeV and 
${\cal D} \equiv (|\Delta \phi_{\vec{\slashed{p}}_T, j_1}|+|\Delta \phi_{\vec{\slashed{p}}_T, j_1}|)/2> \pi/2$. 
In the case of CDF, the cuts applied include jets with $p_T^j>$ 15 GeV and  $|\eta_j|<$ 2.4, 
with the leading and subleading jets required to have $p_T> 25, 20$~GeV, $|\eta_j|<$2 and 
$\Delta R_{jj}>$0.8, and at least one having $|\eta_j|<$0.9. In addition to these cuts, 
we apply the background rejection cuts $\slashed{E}_T > $ 35 GeV, $\Delta \phi (\vec{\slashed{E}}_T, E_T^{j_1})\geqslant 1.5$ 
and  $\Delta \phi (\vec{\slashed{E}}_T, E_T^{j_{2,3}})\geqslant$ 0.4. 

\begin{figure}[h!]
\centering
\includegraphics[scale=0.35]{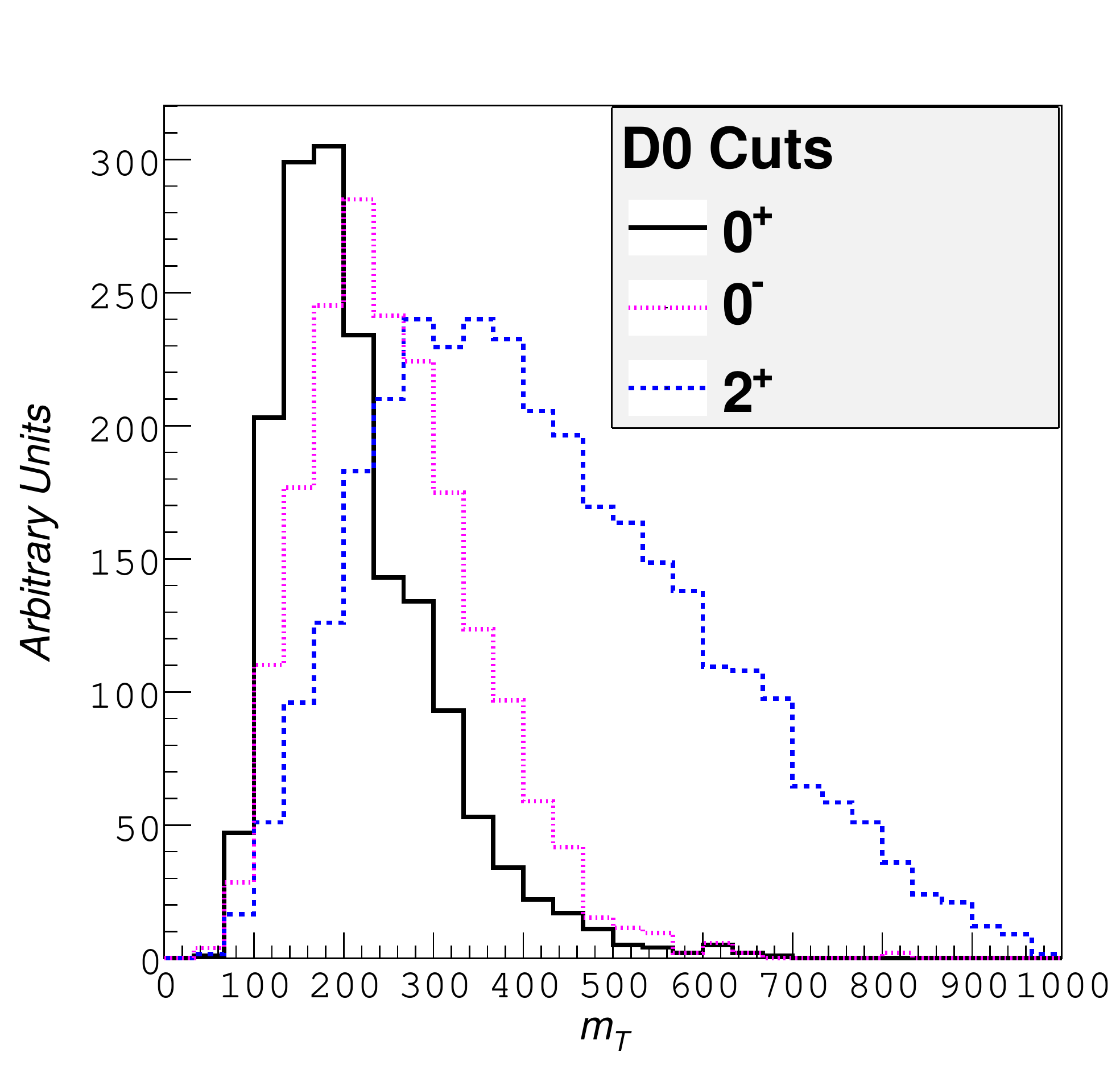}
\includegraphics[scale=0.35]{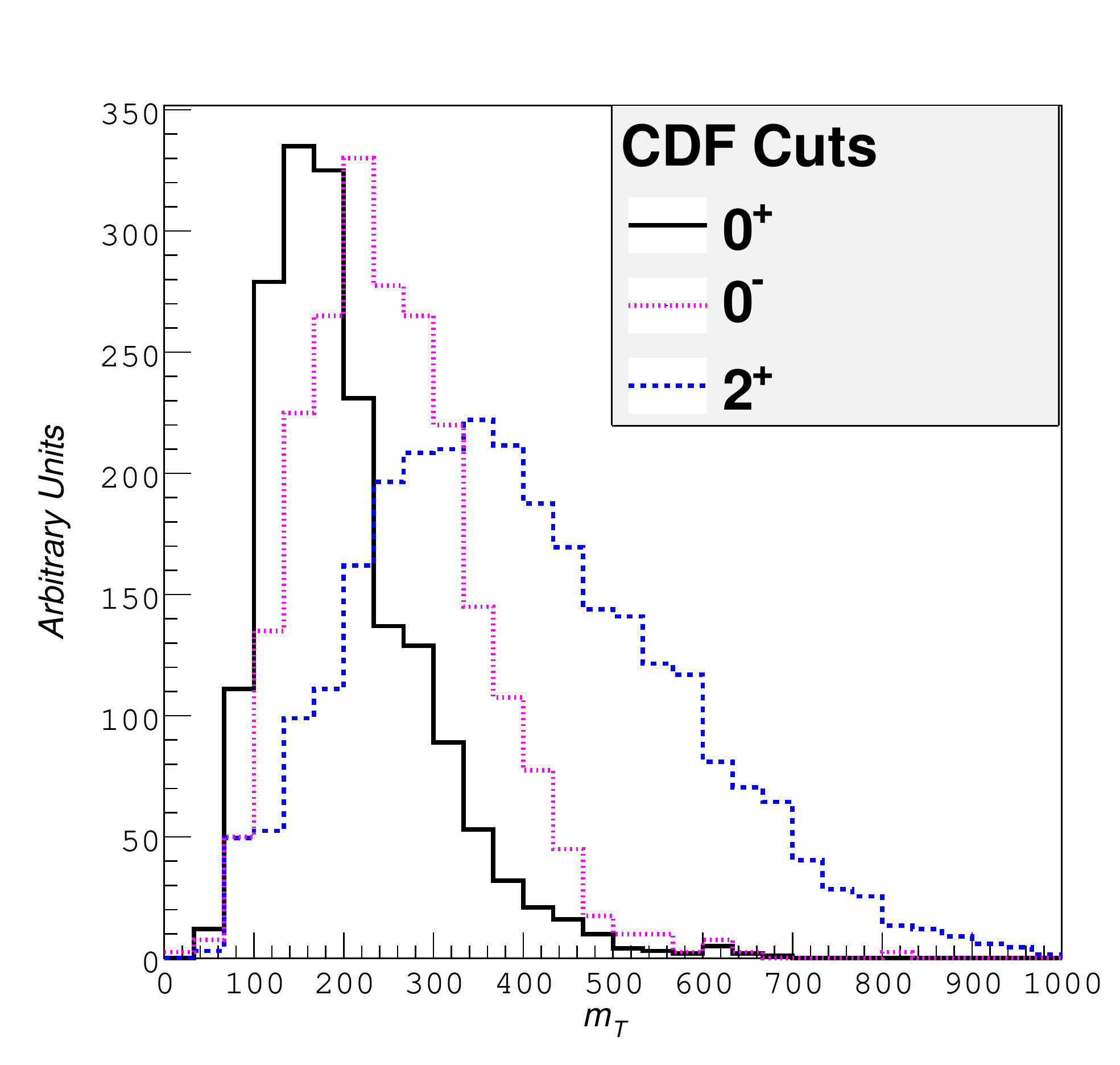}
\caption{\it The effect of a fast simulation of the CDF~\cite{CDF-0l} and D0~\cite{D0-0l} $Z \to {\bar \nu} \nu + X \to \bar{b} b$ analyses
with {\tt Delphes} on the transverse mass distributions for the different $J^P$ assignments.
The discrimination between the different spin-parity assignments seen in Fig.~\ref{fig:beta1or3} is maintained.}
\label{0l}
\end{figure}

We plot in Fig.~\ref{0l} the distributions in the transverse mass variable
\bea
m_T^2=(\slashed{E}_T+E_T^X)^2- (\vec{\slashed{p}}_T+\vec{p}_T^X)^2 ,
\eea
where $X$ corresponds to the two leading jet system.
The number of signal events expected in all the D0 categories quoted as $59 \pm 3$~\cite{D0-0l},
whereas the number of events expected in the CDF analysis is 37, with no errors quoted in~\cite{CDF-0l}.

%%%%%%%%%%%%%
\subsection{LHC $V + X \to \bar{b} b$ analyses}
%%%%%%%%%%%%%%

Both ATLAS~\cite{ATLASVX} and CMS~\cite{CMSVX} have published the results of searches for associated $V + X$ production,
so far establishing upper limits in the absence of a significant signal. 

In simulating the ATLAS analysis with zero leptons, the parton-level cuts we use in our sample generation are $\slashed{E}_T> 120$~GeV, 
$p_T>80$~GeV for the leading jet, and $p_T^j> 20$~GeV for all other jets. We also use the cuts $\Delta \phi_{\, \slashed{E}_T,j}<\pi/2$ 
for the two leading jets. We follow the CMS analysis by including a selection for $V$ and $X$ decays with dijet pairs and 
$V$ decays boosted in the transverse direction, via the cuts listed in Table~\ref{CMSsel}. Other cuts on combinations
such as $m_{jj}$ and $m_{\ell\ell}$ are automatically 100\% efficient for the signal, as is the
requirement for $b$-tagged jets. We note that jets are reconstructed using the anti-$k_T$
algorithm with cut parameter 0.5.

\begin{table}[h!]
\begin{tabular}{|c||c|c|c|}
\hline
 Variable & $W(\ell \nu) X$ &  $Z(\ell \ell) X$  &  $Z(\nu \nu) X$ \\ \hline 
$p_T^{j_1}$ & $>$ 30 GeV & $>$ 20 GeV  & $>$ 80 GeV \\ \hline
$p_T^{j_2}$ & $>$ 30 GeV & $>$ 20 GeV  & $>$ 20 GeV  \\ \hline
$p_T^{jj}$ & $>$ 120 GeV &  --  &  $>$ 120 GeV \\ \hline
$p_T^{V}$ & $>$ 120 GeV &  $>$ 50 GeV & --  \\ \hline
$\Delta \phi_{\, \slashed{E}_T,j}$ & --  & -- & 0.5 \\ \hline
$\slashed{E}_T$ & $>$ 35 GeV (e) &  -- & $>$ 120 GeV \\ \hline
\end{tabular}
\caption{\it Cuts used by CMS in their search for associated $V + X$ production. Note the
selections for dijet pairs and $V$ decays boosted in the transverse direction.}
\label{CMSsel}
\end{table}

\begin{figure}[h!]
\centering
\includegraphics[scale=0.27]{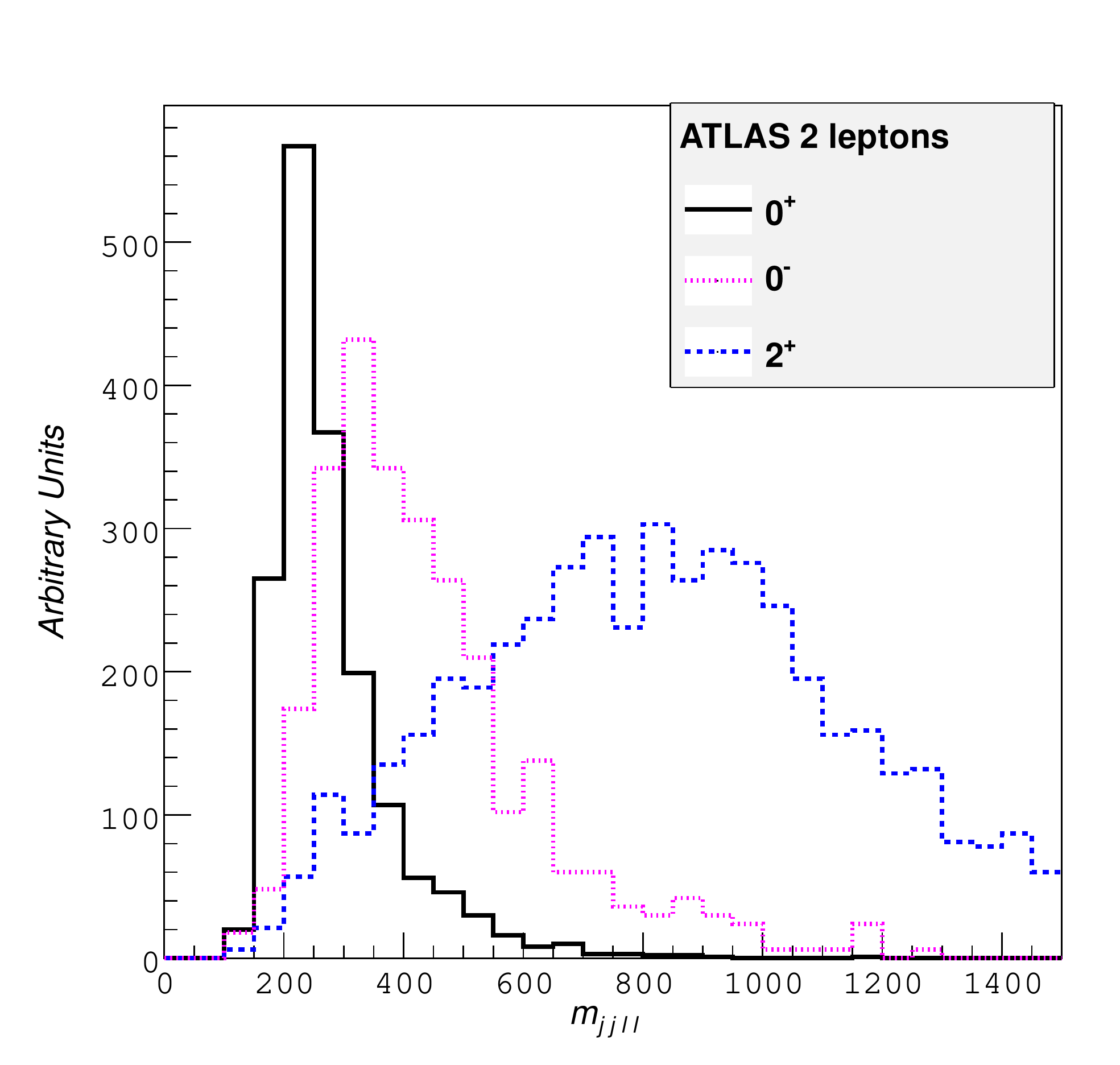}
\includegraphics[scale=0.27]{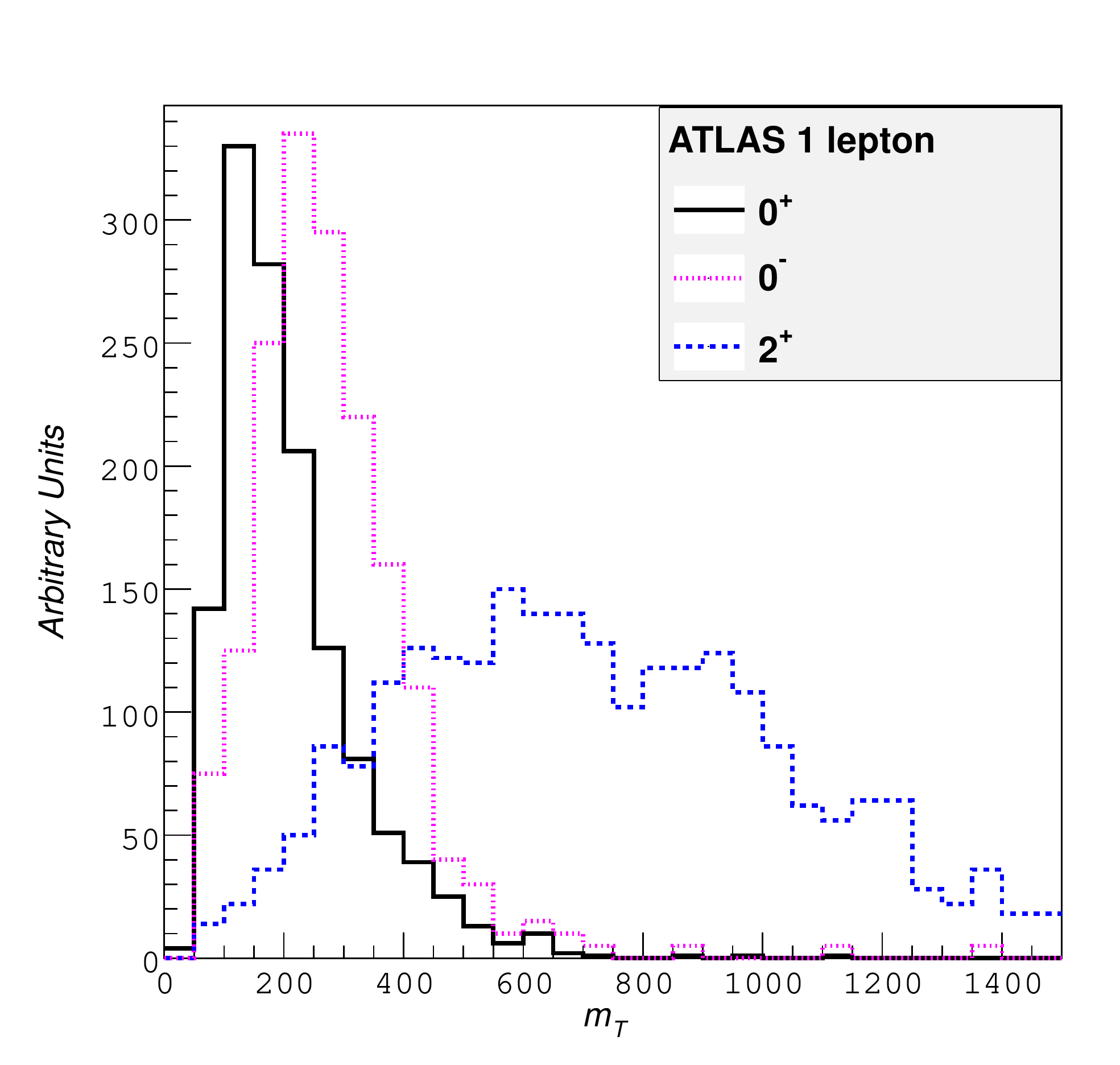}
\includegraphics[scale=0.27]{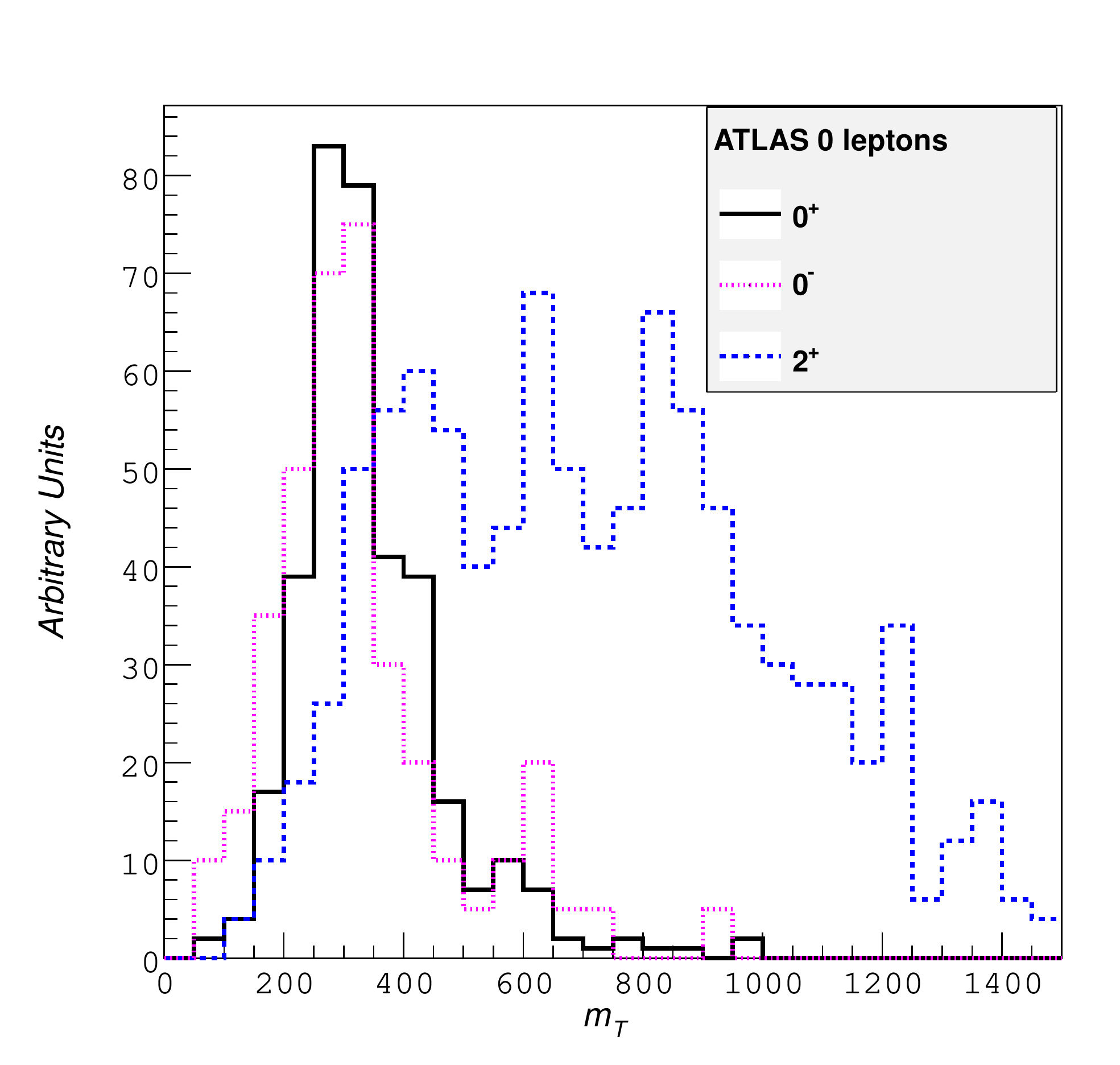}
\includegraphics[scale=0.27]{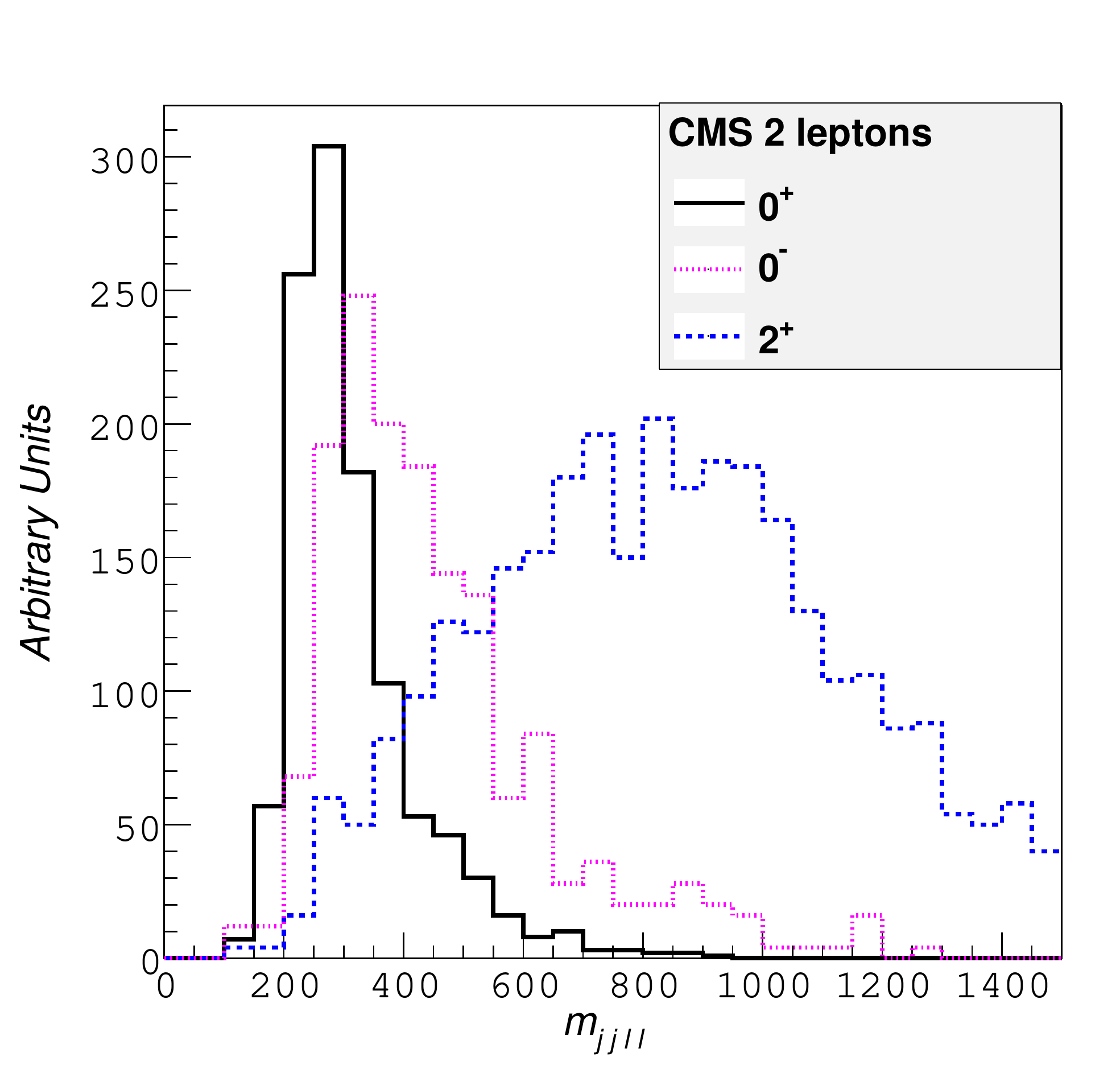}
\includegraphics[scale=0.27]{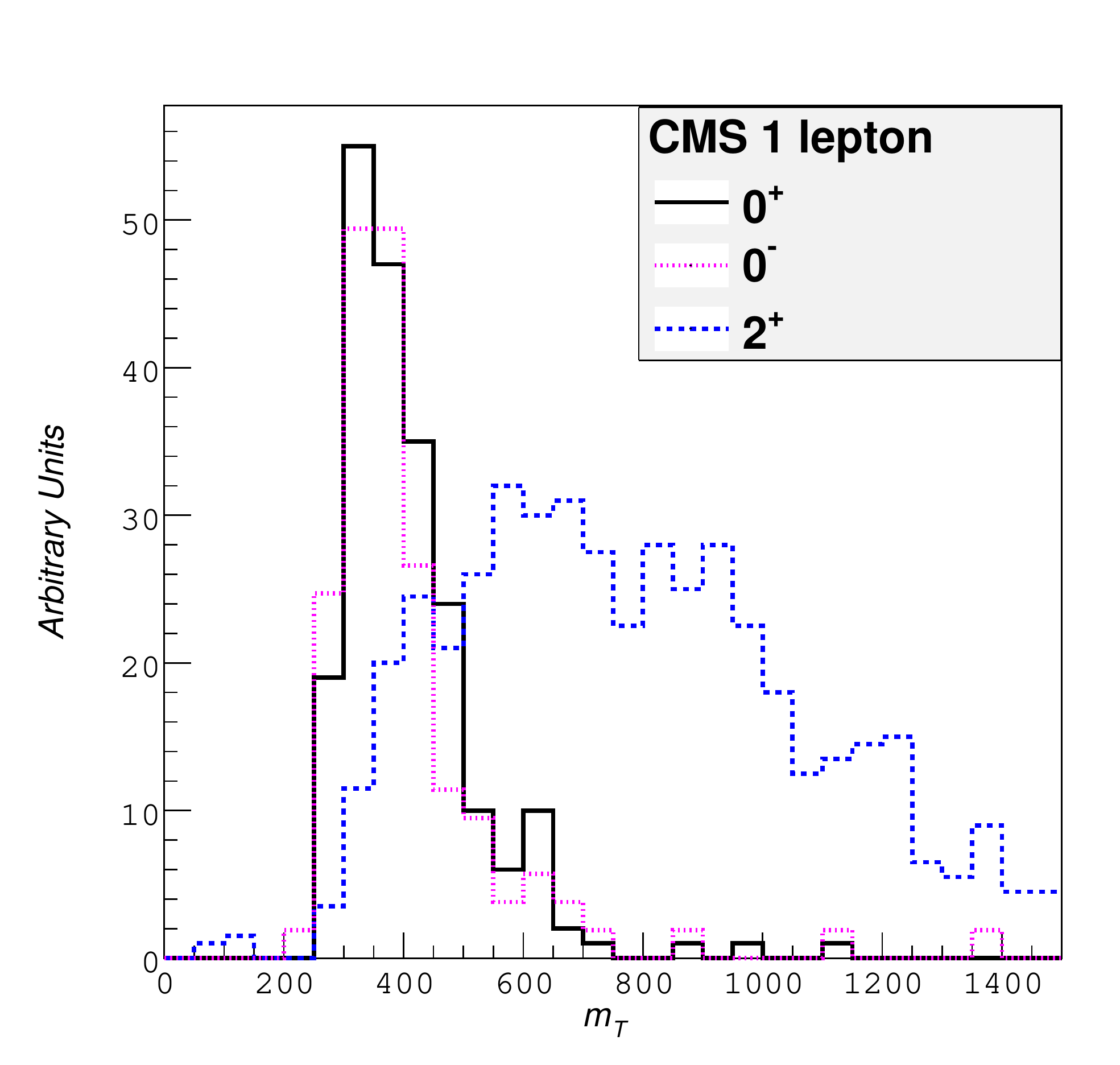}
\includegraphics[scale=0.27]{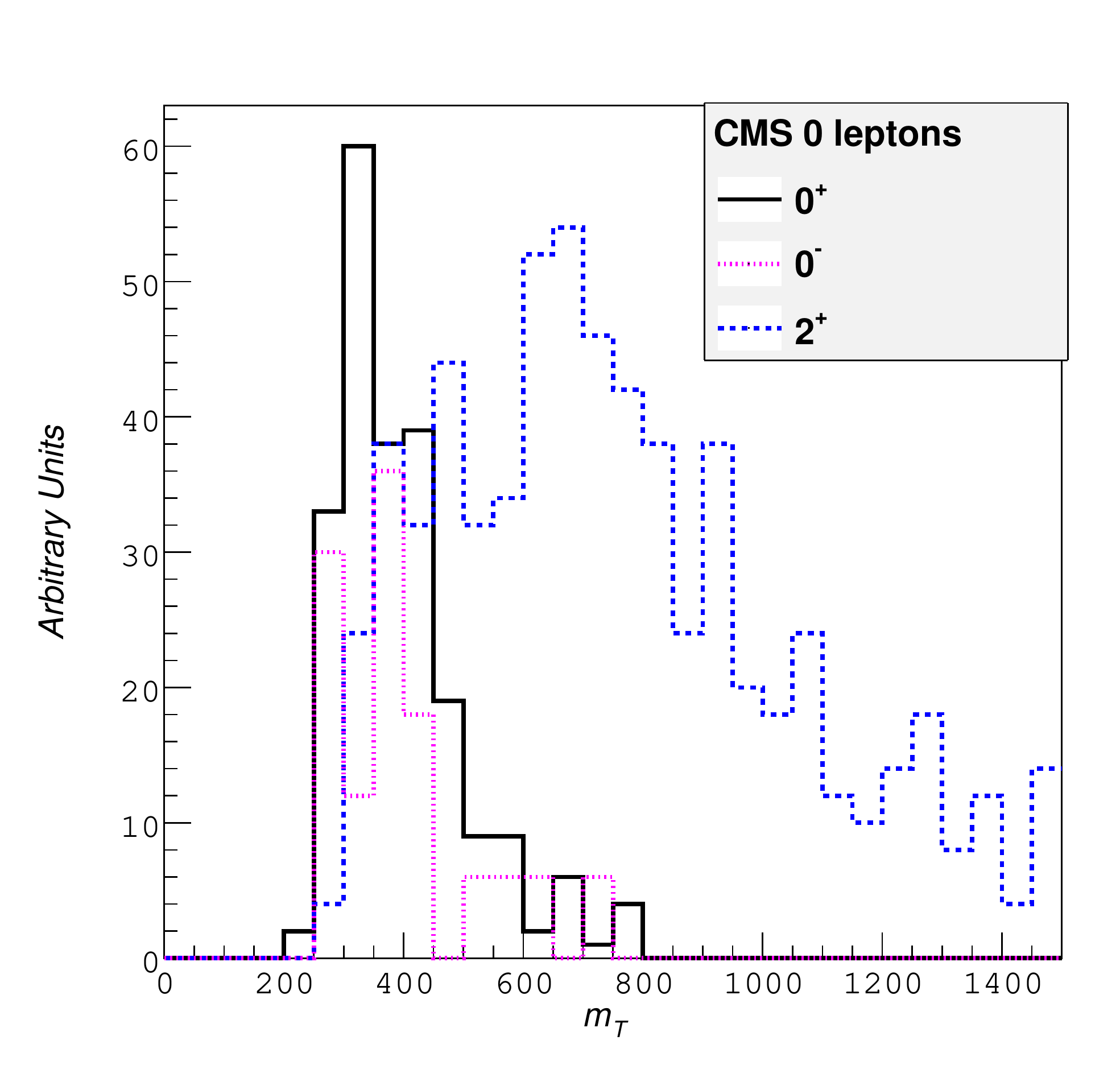}
\caption{\it Kinematical distributions for the 7-TeV ATLAS (upper row) and CMS (lower row) $V + X \to \bar{b} b$ analyses
in the two-, one- and zero-lepton cases (left, centre and right panels, respectively).}
\label{ATLAS}
\end{figure}

We display in Fig.~\ref{ATLAS} various kinematical distributions found after simulations of
the ATLAS cuts (upper row) and the CMS cuts detailed in Table~\ref{CMSsel} (lower row)
for events with two, one and zero identified leptons (left, centre and right panels). In almost every
case, we see that the distributions for the $0^+, 2^+$ and $0^-$ spin-parity assignments
for the $X$ particle are clearly distinguishable. The only exceptions are provided by the
transverse mass distributions for the CMS analysis of one- and zero-lepton events, where
we see that the $0^-$ and $0^+$ cases are indistinguishable. This is a consequence of
the boost requirements, which suppress low-mass $V + X$ combinations. These
requirements also squeeze together the $m_{VX}$ distributions for the two-lepton
$0^+$ and $0^-$, though these are still distinguishable in principle.

%%%%%%%%%%%%%%%%%
\section{Statistical Procedure}
%%%%%%%%%%%%%%%%%%%

The kinematic variable of interest for our analysis is $x \equiv M_{VX}$ or the related quantity $M_T$, 
and we can quantify the significance of the 
separation between different spin-parity hypothesis through the use of a likelihood for the distribution in $x$. 
Since we are dealing with low statistics we consider an unbinned likelihood in the spirit of~\cite{Cousins}. 

The likelihood of a single event $x_i$, for a spin-parity hypothesis $s = 0^+, 0^-,2^+$, is given by a probability 
density function $\text{pdf}_s(x_i)$. This pdf is a normalized, high-statistics Monte Carlo histogram
that takes into account detector acceptance effects and cuts on the distribution of the kinematic variable $x$. 
The full likelihood for $x$ is then obtained by multiplying the pdf for each event $i$:  
\begin{equation}
	\mathcal{L}_s = \prod_{i=1}^M \text{pdf}_s(x_i)		\quad .
\label{pdf}
\end{equation}
We follow the Neyman-Pearson approach in using the log-likelihood ratio for our test statistic, defined as 
\begin{equation}
	\Lambda = -2\ln \left(\frac{\mathcal{L}_A}{\mathcal{L}_B}\right)		\quad .
\label{calL}
\end{equation}
The separation significance between two spin-parity hypotheses A and B can be estimated by generating a 
large number of toy experiments to obtain a distribution in $\Lambda$. If the toys are generated for hypothesis A, 
then the distribution of $\Lambda$, $f_A(\Lambda)$, will be centered around a negative mean value. 
Conversely, for toys generated according to hypothesis B the $\Lambda$ distribution $f_B(\Lambda)$
will be centered around a positive mean, with the tails of the two distributions $f_A$ and $f_B$ overlapping to a certain extent.

For a given observed $\Lambda_\text{obs}$, the probability of getting a more extreme value of $\Lambda$ 
than the one observed assuming hypothesis A is
\begin{equation}
	\alpha = \left\{ 
		\begin{array}{l} 
			\frac{1}{N}\int_{\Lambda_\text{obs}}^\infty f_A(\Lambda) d\Lambda	\quad \Lambda_\text{obs} \geq \Lambda_\text{mean} \\
			\\
			\frac{1}{N}\int^{\Lambda_\text{obs}}_{-\infty} f_A(\Lambda) d\Lambda	\quad \Lambda_\text{obs} < \Lambda_\text{mean}
		\end{array}
		\right.
\label{alpha}
\end{equation}
where $N = \int_{-\infty}^\infty f_A(\Lambda) d\Lambda$. A similar definition can be given for the probability $\beta$
assuming hypothesis B instead. These can be identified with the p-values quantifying the agreement between the 
observed data and the hypotheses. 

We may restrict the definition of  $\alpha$ and $\beta$ to be always the integral towards the right and left tail 
end of the distribution, respectively. Then $\alpha$ is also defined as the ``type I'' error, namely
the probability of rejecting hypothesis A given that it is true, and $\beta$ is the ``type II'' error, 
namely the probability of wrongly accepting hypothesis A given that B is actually true. 
The ``power'' of the test is $1-\beta$, so that a high probability of getting a type II error corresponds to a test with weak power. 

There are two ways of reporting the expected significance, reflecting different underlying philosophies. 
The first takes an asymmetric approach to the two hypotheses: the mean value of $\Lambda$ under
hypothesis A is the value of $\Lambda_\text{obs}$ that an experiment is expected to measure if hypothesis A is true,
and one may quote the p-value $\beta$, the level at which we will then be able to exclude hypothesis B. 
By randomly sampling $\Lambda_\text{obs}$ from $f_A$, one can give one-sigma bands for the expected 
significance for $\beta$. In this approach the value of $\alpha$ and $\beta$ defined as the acceptance limit is 
fixed (for example to 0.05) and we seek to minimize $\beta$, the type II error. The second approach instead 
treats the two hypotheses equally by defining the acceptance region for hypothesis A (B) as lying to the left (
right) of $\Lambda_\text{cutoff}$ respectively, where $\Lambda_\text{cutoff}$ is the value of $\Lambda$ for 
which $\alpha = \beta$. Thus, whatever the value of $\Lambda_\text{obs}$, the significance with which one 
hypothesis can be considered excluded and the other accepted is $\alpha$ $(=\beta)$. 

It is clear from these two definitions of expected significance that, given a distribution of $\Lambda$ for the two hypotheses, 
the second (symmetric) approach will yield a more conservative significance than the first (asymmetric method). Since it is also the more objective method, below we quote this symmetric approach for the significance. 

The significance $\alpha$ is usually translated into $n$ standard deviations by finding the equivalent area 
under a standard Gaussian distribution~\footnote{This is the one-sided definition most commonly used in 
the literature, as opposed to the two-sided convention sometimes seen, which generally
yields a higher number of standard deviations for the same p-value.}:
\begin{equation}
	\alpha = \frac{1}{\sqrt{2\pi}} \int_n^\infty e^{-\frac{x^2}{2}} dx	\quad . 
\label{alpha2}
\end{equation}
For example, $\alpha=0.05$ corresponds to $n=1.64$, and the discovery standard of $n=5$
corresponds to $\alpha = 2.87 \times 10^{-7}$. 

%%%%%%%%%%%%%%%%%
\section{Analysis using `toy' experiments}
%%%%%%%%%%%%%%%%%%%

We evaluated the expected separation significance using both the symmetric and the asymmetric 
method~\footnote{Though below we quote results only for the former, more conservative, approach.},
by generating 100 `toy' experiments corresponding to each of the analyses discussed above,
namely the CDF, ATLAS and CMS 0-, 1- and 2-lepton analyses, and the D0 0- and 2-lepton analyses.
These toys are designed to reproduce the statistics found in the corresponding analyses after
implementing the event selections and cuts. We have checked in specific cases that the
separation significances quoted below are quite insensitive to the number of `toys'
beyond 100. In modelling each analysis, 
we neglect the contaminations by backgrounds: their simulation would be more
complicated and take us beyond the scope of this work. We note that the backgrounds in the
TeVatron analyses are in any case very small in the bins with log$_{10} (s/b) > - 1.5$~\cite{CDFD0}. The
backgrounds in the LHC analyses are currently larger, but we expect them to decrease as the
analyses are refined.

\begin{figure}[h!]
\centering
\includegraphics[scale=0.35]{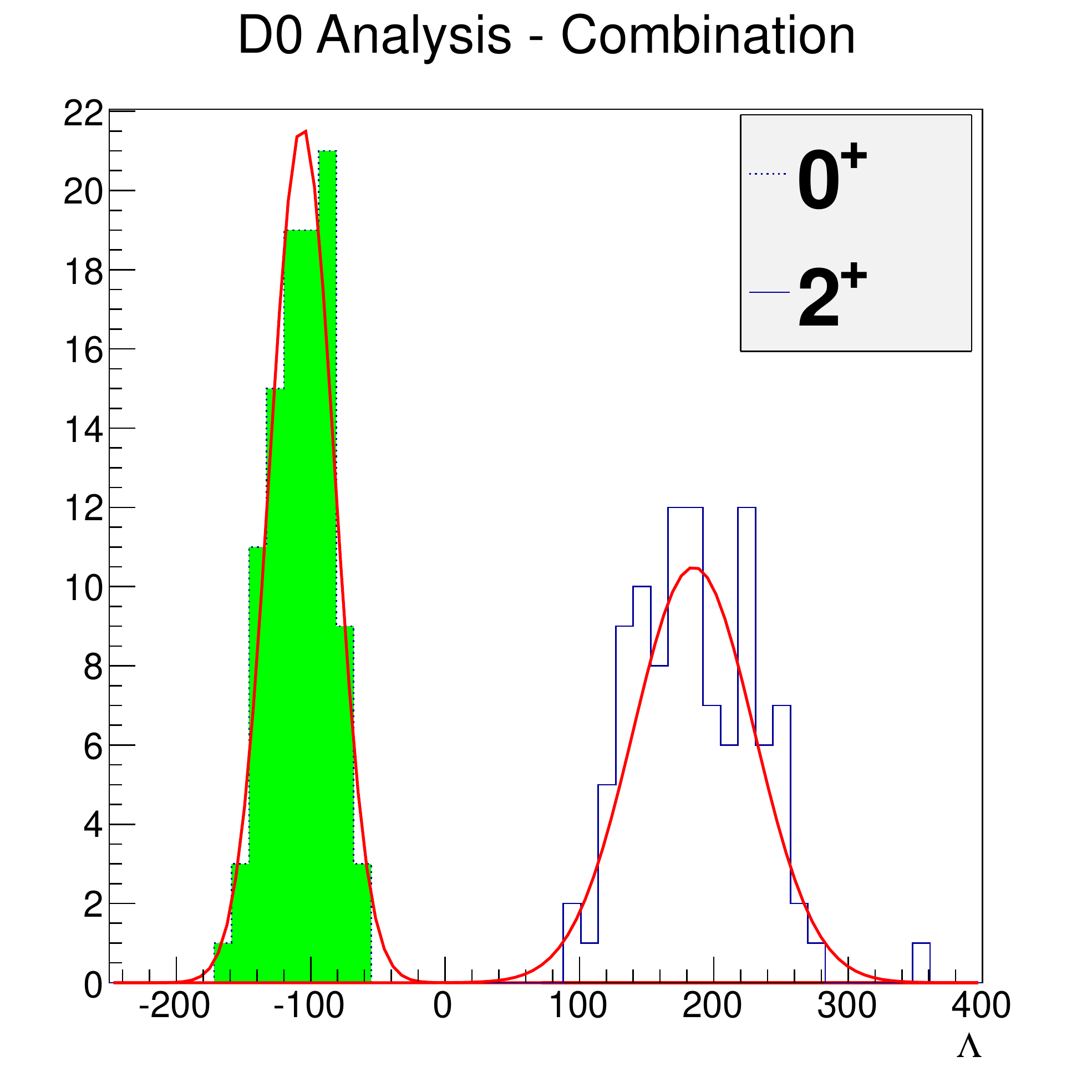}
\caption{\it An example of the statistical separation that could be achieved between the spin-parity
assignments $J^P = 0^+$ and $2^+$. It is based on a combination of sets of 100 `toy' experiments simulating the D0
zero-, one- and two-lepton analysis separately, each with a number of events surviving the experimental cuts that is
similar to that found in the D0 analysis.}
\label{toyplot}
\end{figure}

Fig.~\ref{toyplot} illustrates how these toys can be used to estimate the statistical separations 
between a pair of $J^P$ hypotheses that can be achieved, using the example
of the D0 zero-, one- and two-lepton analyses. A set of 100 `toy' experiments was generated for
each of these analyses, and the results combined. The horizontal axis is the symmetric test statistic
$\Lambda \equiv -2 {\rm ln}({\cal L}_A/{\cal L}_B)$ (\ref{calL}). The separation between the distributions generated for the
$J^P = 0^+$ and $2^+$ hypotheses (green shaded dotted blue and open blue solid 
histograms, respectively) is clear.

Table~\ref{table:separationsignificance} summarizes the statistical separations we find for each TeVatron
and LHC analysis between the $0^+$ and $2^+$ hypotheses and between the $0^+$ and $0^-$ hypotheses (in parentheses).
As one would expect from the invariant-mass distributions shown earlier, the
statistical separation between the $0^+$ and $2^+$ hypotheses is generally stronger than that
between the $0^+$ and $0^-$ hypotheses.
Also shown are the separation significances we find for the combinations of analyses in the TeVatron experiments.
We note that in these experiments the approximate overall significance of
their evidence for $X$ production in association with vector bosons~\cite{CDFD0} $\sim 3 \sigma$,
so results above this level for the separation significance are only formal. In the cases of the LHC
experiments, we quote results only for the two-lepton analyses, as the separation significances we find
in their one- and zero-lepton analyses are much lower. Since the backgrounds in the LHC experiments are
currently large compared with any signals, our results (which assume negligible backgrounds)
are not directly applicable at present. However, we expect the signal/background ratios to increase as the analyses
progress, and the results in Table~\ref{table:separationsignificance} suggest levels of separation
to which improved analyses could aspire.

\begin{table}[h!]
	\center
	\begin{tabular}{ | c | c | c | c | c |}
		\hline
		 Experiment & Category & Hypothesis A & Hypothesis B & Significance in $\sigma$ \\ \hline
		CDF & 0l & $0^+$ & $2^+ (0^-)$ & 3.7 (1.3)    \\ \hline
		%CDF & 0l & $0^+$ & $0^-$  & 4.4 \\ \hline
		 & 1l & $0^+$ & $2^+ (0^-)$ &  2.5 (1.0)    \\ \hline
		%CDF & 1l & $0^+$ & $0^-$ &    \\ \hline
		 & 2l & $0^+$ & $2^+ (0^-)$ & 1.4 (0.78)   \\ \hline
		%CDF & 2l & $0^+$ & $0^-$ &  \\ \hline
		 & ~Combined~ & $0^+$ & $2^+ (0^-)$ & 4.8 (1.6)   \\ \hline\hline
		%CDF & Combined & $0^+$ & $0^-$ &    \\ \hline\hline
		D0 & 0l & $0^+$ & $2^+ (0^-)$ & 3.5 (1.2)   \\ \hline
		%D0 & 0l & $0^+$ & $0^-$ & 4.2    \\ \hline
		 & 2l & $0^+$ & $2^+ (0^-)$ & 1.8 (1.2)     \\ \hline
		%D0 & 2l & $0^+$ & $0^-$ &    \\ \hline\hline
		& Combined & $0^+$ & $2^+ (0^-)$ & 4.0 (1.6)  \\ \hline\hline
		%CDF & Combined & $0^+$ & $0^-$ & ?   \\ \hline\hline
		%ATLAS & 0l & $0^+$ & $2^+/0^-$ & ?      \\ \hline
		%ATLAS & 0l & $0^+$ & $0^-$  & ? \\ \hline
		 %& 1l & $0^+$ & $2^+/0^-$ &  ?      \\ \hline
		%ATLAS & 1l & $0^+$ & $0^-$ & ?    \\ \hline
		 ATLAS & 2l & $0^+$ & $2^+ (0^-)$ & 2.4 (1.1)    \\ \hline
		CMS & 2l & $0^+$ & $2^+ (0^-)$ & 2.3 (0.70)     \\ \hline
	\end{tabular}
	\caption{\it The separation significances between different $J^P$
	hypotheses estimated for each Tevatron and LHC experiment, using in each case 100 `toy' experiments with
	similar event numbers to the data, using the symmetric method of hypothesis testing described in the text.\ }
	\label{table:separationsignificance}
\end{table}

\begin{figure}[htb!]
\centering
\includegraphics[scale=0.8]{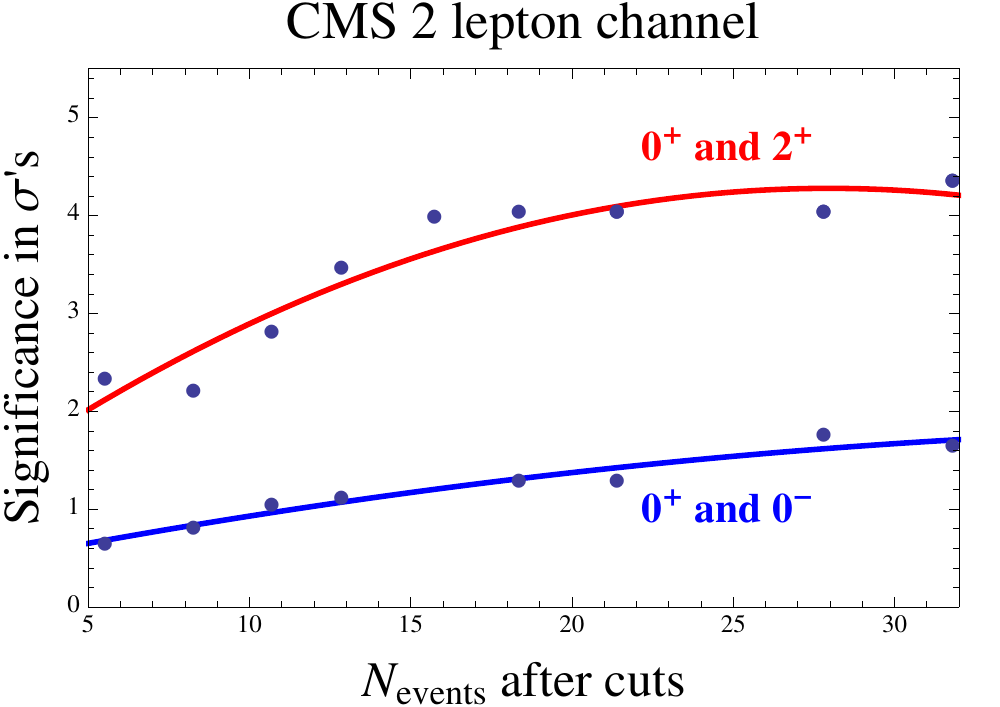}
\includegraphics[scale=0.8]{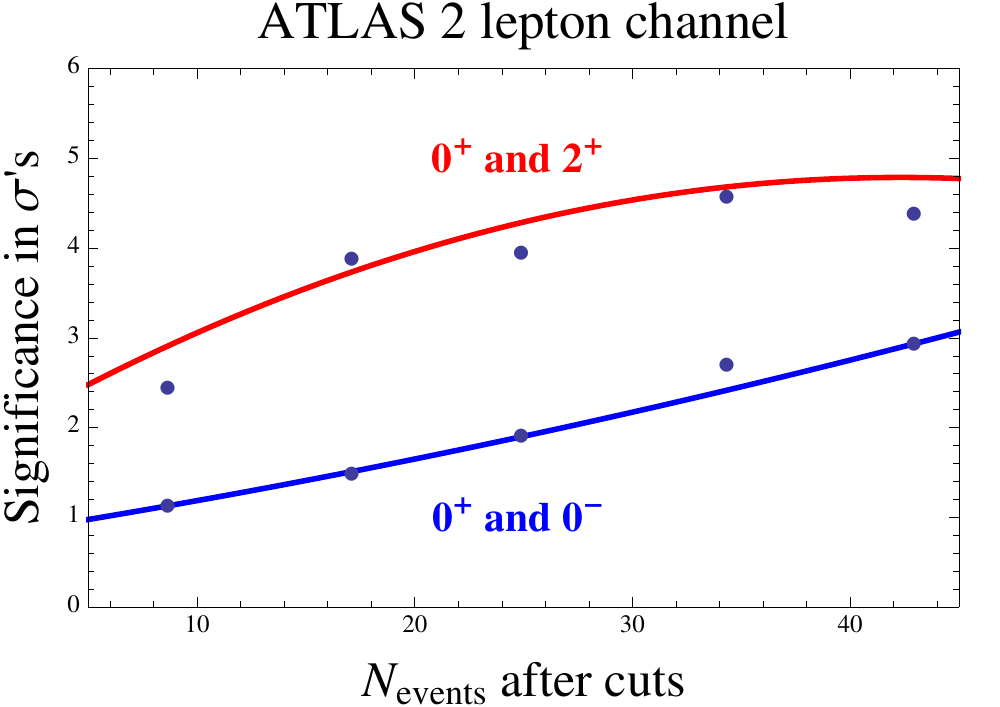}
\caption{\it The variation in the significance of the statistical separation obtainable as a
function of the number $N$ of events surviving the CMS (left) and ATLAS (right) experimental 
event selections and cuts for two-lepton events,
for distinguishing between $0^+$ and $2^+$ (upper points and red lines)
and between $0^+$ and $0^-$ (lower points and blue lines).}
\label{fig:largertoys}
\end{figure}

In the cases of the LHC analyses, we have also generated toys simulating the larger numbers
of signal events that will become available in the future, with the results illustrated in 
Fig.~\ref{fig:largertoys} (neglecting backgrounds, as before). The left panel shows how the
statistical significance in the CMS two-lepton analysis, in numbers of $\sigma$, 
of the separations between the $0^+$ and $2^+$ hypotheses 
(upper points and red line to guide the eye) and between the $0^+$ and $0^-$ hypotheses (lower points and blue line) would
increase with the number of signal events $N$ surviving the experimental selection and cuts.
The right panel shows a similar analysis for the ATLAS two-lepton analysis. 
We see again that it will be easier to discriminate between the
$0^+$ and $2^+$ hypotheses than between the $0^+$ and $0^-$ hypotheses,
and that the hypotheses can be distinguished cleanly if the backgrounds can be suppressed.

%%%%%%%%%%%%%%%%%
\section{Conclusions}
%%%%%%%%%%%%%%%%%%%

We have shown in this paper that the invariant mass distributions for $V + X$ combinations, $M_{VX}$, 
are theoretically very different for the $J^P = 0^+, 0^-$ and $2^+$ assignments for the new boson with
mass $\sim 125$~GeV recently discovered by the ATLAS and CMS Collaborations~\cite{ATLAS,CMS}.
Making simulations using {\tt PYTHIA} and {\tt Delphes}, we have also shown that these differences
survive the experimental event selections and cuts in searches for $X$ production in association with
two-, one- and zero-lepton decays of the heavy vector bosons $Z$ and $W$. We have also used
simulated `toy' experiments to estimate the statistical separations that could in principle be attained
by the CDF, D0, ATLAS and CMS experiments if the experimental backgrounds were negligible.

In the case of the TeVatron experiments, our analysis indicates that the data currently available
should be able to discriminate between the $0^+$ and either the $2^+$ or $0^-$ hypotheses with
high significance, assuming that the backgrounds are small. The latter is not a good assumption
for the LHC experiments, but we show what statistical separations might be attainable with
increased data sets and reduced backgrounds.

Analyses of the possible backgrounds go beyond the scope of this paper. However, we
think that our analysis already demonstrates the potential of $M_{VX}$ measurements
to provide valuable insight into the possible $J^P$ assignment of the $X$ particle.
It may well be that its $J^P$ will be determined by a combination of different measurements
that each make contributions to the global likelihood. In this perspective, we hope that the
the $M_{VX}$ measurements proposed here will play useful roles.

\section*{Acknowledgements}
We thank Ricky Fok for valuable discussions on related subjects.
The work of JE and TY was supported partly by the London
Centre for Terauniverse Studies (LCTS), using funding from the European
Research Council via the Advanced Investigator Grant 267352.
The work of DSH was supported partly by Korea Foundation for International Cooperation of 
Science \& Technology (KICOS) and Basic Science Research Programme through the 
National Research Foundation of Korea (2012-0002959).
The authors thank CERN for kind hospitality, and TY additionally thanks
Prof. T.~Kobayashi and the Bilateral International Exchange Program of Kyoto University for
kind hospitality during the completion of this work.

%\newpage

%%%%%%%%%%%%%%%%%%%%%%%%%%
%%%%%%%%%%%%%%%%%%%%%%%%%%%%%%%%%%%%%%%%%%

\end{document}